\documentclass{article}
\usepackage{arxiv}                  
\usepackage[utf8]{inputenc}         
\usepackage[T1]{fontenc}            
\usepackage{textcomp}              
\usepackage{lmodern}                
\usepackage{amsmath}                
\usepackage{amssymb}                
\usepackage{graphicx}               
\usepackage[caption=false]{subfig}  
\usepackage{natbib}                 
\usepackage{hyperref}               
\usepackage{url}                    

\title{Atmospheric Predictability Beyond 30 Days with Machine Learning}

\author{%
  P. Trent Vonich\thanks{Corresponding author: P. Trent Vonich, tvonich@uw.edu} \\
  University of Washington \\
  Air Force Institute of Technology \\
  \And
  Gregory J. Hakim \\
  University of Washington \\
}
\newcommand{\absbody}{Atmospheric predictability research has long held that rapid error growth at small spatial scales imposes an intrinsic limit of roughly two weeks on deterministic weather forecast skill. We challenge this limit using GraphCast, a machine-learning weather model, by optimizing initial conditions for twice-daily forecasts spanning 2020. This approach yields an average error reduction of 86\% at ten days relative to control forecasts from reanalysis initial conditions, with skill lasting beyond 30 days. Mean optimal initial-condition perturbations reveal large-scale, spatially coherent corrections primarily reflecting an intensification of the Hadley circulation. Forecasts using GraphCast-optimal initial conditions in the Pangu-Weather model achieve a 21\% error reduction, peaking at four days, indicating that analysis corrections reflect adjustments that target both model and analysis error. These results demonstrate the existence of initial conditions producing skillful deterministic forecasts far beyond two weeks. Whether such initial conditions can be identified in real-time for improving operational weather forecasts remains a topic of future research.} 

\begin{document}

\maketitle

\begin{abstract}
\absbody
\end{abstract}

%
%
%
\section*{Significance Statement}
     \begin{itemize}
         \item Our results contest the long-standing view that rapid error growth at small spatial scales fundamentally limits reliable weather forecasts to about two weeks. By applying machine-learning based optimization to atmospheric initial conditions, we achieve skillful deterministic hindcasts to at least 33 days---more than twice the conventional limit. Validating these optimized initial conditions in an independent weather model leads to significant but smaller improvements, highlighting the importance of model error in the procedure. On average, the optimal corrections reveal physically consistent large-scale patterns that align with an intensified tropical overturning circulation. These results demonstrate the existence of initial conditions that considerably extend the current established limit of predictability. Real-time determination of such initial conditions for operational forecasting remains an open question.
     \end{itemize}
%

%

%
\section{Introduction}

For more than half a century, atmospheric predictability has been framed by Edward Lorenz’s seminal concept of the ``butterfly effect,'' which proposes that infinitesimal errors in initial conditions grow rapidly, ultimately limiting skillful deterministic weather forecasts to approximately two weeks \citep{Lorenz1969}. This paradigm has profoundly shaped meteorological science, fostering the prevailing view that chaos imposes an insurmountable boundary on weather forecasting in the absence of other sources of skill (e.g., the ocean). 

Although frequently linked to Lorenz, the two-week predictability limit actually originates from \citet{Charney1966}, which reported a 5-day doubling time of errors in a first-generation general circulation model. Extrapolation of these findings suggested that the intrinsic predictability limit for Earth's atmosphere was about two weeks. This view has widely influenced scientific and public expectations of weather model performance. However, a series of recent papers \citep{Shen2022,Shen2023,Shen2024} clarify that while Lorenz's original 1969 model is an effective illustration of chaotic dynamics, it is ill-suited for quantifying the atmosphere’s intrinsic predictability due to its absence of baroclinic and dissipative processes---a critique Lorenz would later acknowledge himself \citep{lorenz1996predictability}.

Modern experiments utilizing models that do include these processes increase the intrinsic limit modestly beyond two weeks. For example, \citet{zhang2019predictability} find that reduction of current-day operational forecasting initial condition error by an order of magnitude would extend mid-latitude skill to about 15 days. Similarly, ``perfect twin'' experiments with convection-allowing models show slightly longer limits, with errors plateauing at 17 days in the mid-latitudes and beyond 20 days in the tropics \citep{judt2018, judt2020, selz2019}. Ensemble forecasts average over random errors and therefore have extended skill up to approximately 23 days \citep{buizza2015forecast}.

The emergence of machine-learning (ML) weather models provides a new tool to assess predictability and reduce errors by adjusting forecast initial conditions. A case study demonstration \citep{heatwave_2024} uses backpropagation and gradient descent techniques to create an optimal initial condition, defined as the input that best reproduces a target sequence. With full knowledge of the future trajectory, the deterministic method finds the initial condition that produces a more accurate forecast without an ensemble. This method resembles classical adjoint sensitivity approaches \citep[e.g.,][]{langland1995,langland2002,doyle2012,doyle2014,doyle2019,lloveras2025_goose}, except that the full nonlinear ML model operates substantially faster than traditional weather forecasting models \citep{heatwave_2024, bano_medina2025}. Applying this technique to the June 2021 Pacific Northwest heatwave \citep{thompson20222021,leach2024heatwave} using the GraphCast model \citep{lam2023learning}, \citet{heatwave_2024} show that the optimized forecast achieves an 85\% reduction in 10-day error compared to a control originating from an ECMWF Reanalysis Version 5 (ERA5) initial condition \citep{hersbach2020era5}, with improvements decreasing to zero around 22.5 days. Moreover, forecasts with a different model (Pangu-Weather; \citealt{bi2023}) initialized with the GraphCast-optimized inputs show comparable 10-day forecast improvements, suggesting that model error is not a critical component of the optimal initial condition. 

Here we increase the sample size to address three questions:
\begin{enumerate}
    \item How consistently does initial condition optimization enhance forecast accuracy?
    \item What is the maximum lead time for which forecast skill can be achieved with this approach?
    \item How reliably can optimized initial conditions produced by GraphCast improve predictions in a different model?
\end{enumerate}

Unlike classical predictability studies, which define a limit by the divergence of nearby states or by ensemble spread approaching climatology, we define predictability as the lead time beyond which adjustments to the initial condition no longer reduce forecast error. This reframing provides an objective definition of the predictability limit for individual trajectories that is independent of error growth rate, amplitude, and representation within a climatological sample. Results presented here provide an existence proof of initial conditions that evolve with sustained accuracy well beyond the conventional two-week limit of predictability.  

We refine the original optimization method of \citet{heatwave_2024} and apply it to 732 unique initialization times, forecasts generated at 00Z and 12Z for every day of 2020, and verify the outputs against ERA5. Results show 10-day forecast improvements that are similar in magnitude to that of the 2021 heatwave study (86\%) and forecast skill that extends to about double the value of current estimates of intrinsic predictability. When tested in Pangu-Weather, the GraphCast-derived optimal initial conditions yield statistically significant, but smaller, improvements, suggesting both genuine reduction of initial-condition error and model-specific bias correction.

\section{Data and Methods}

\subsection*{Model}
We use the ``small'' version of the GraphCast model \citep{lam2023learning}, selected for its modest memory footprint. This enables gradient computations over extended windows, up to 32 days, in a practical timeframe. Since this configuration operates on a 1.0° grid, it does not resolve mesoscale processes. As a result, our findings apply only to predictability at synoptic and planetary scales.

GraphCast forecasts six atmospheric state variables: geopotential, temperature, specific humidity, vertical velocity, and zonal and meridional wind components, resolved across 13 pressure levels. It also predicts four surface variables—mean sea-level pressure, 2-meter air temperature, and 10-meter zonal and meridional wind components—alongside 6-hour accumulated precipitation, all on a 1.0° × 1.0° grid. With 36.7 million parameters, GraphCast was trained on ERA5 reanalysis data from 1979 to 2015 \citep{hersbach2017era5}. None of the forecasts optimized in this study are part of the training data.

To generate predictions, inference utilizes two atmospheric input states, separated by a 6-hour interval, producing a single output state 6 hours in the future. For extended forecasts, the output is autoregressively fed back into the model alongside the prior 6-hour state, enabling indefinite prediction. In this study, optimal perturbations are computed exclusively for the state variables, while static fields, including the land-sea mask and surface geopotential, remain unaltered. 

\subsection*{Optimization} 
Our approach leverages the fully differentiable nature of GraphCast to optimize initial conditions in a nonlinear framework, overcoming the limitations of traditional adjoint models \citep[e.g.,][]{langland1995,errico1997adjoint} related to computational expense. Machine-learning models integrate linear and nonlinear operations across layers, enabling seamless derivative computation via the chain rule. In this study, all automatic differentiation is performed using GraphCast implemented in the JAX framework \citep{lam2023learning}. JAX provides robust support for automatic differentiation, complemented by GPU acceleration and dynamic code optimization \citep{jax2018github}.

This differentiable framework enables iterative refinement of atmospheric initial conditions. Given the input state ${\bf x}_i$ for iteration $i$ (where $i=0$ is the unaltered ERA5 initial condition), we compute an increment based on the gradient of the forecast loss, $\mathcal{L}({\bf N}({\bf x}_i))$, with respect to the inputs. Here, ${\bf N}$ denotes GraphCast autoregressive inference starting from the chosen forecast optimization time. The update for each iteration is defined by

\begin{equation}
{\bf x}_{i+1} = {\bf x}_i - \eta\frac{\partial \mathcal{L}}{\partial {\bf x}_i}
\label{eqn:increment}
\end{equation}

\noindent The derivative in (\ref{eqn:increment}) entails tracing the loss gradient back through the GraphCast neural network for every 6-hour time step. This gradient characterizes how changes to the inputs influence the loss function. Adjoint models can also be employed to perform gradient descent, but, like deep learning models, they may struggle to navigate complex gradient landscapes with numerous saddle points and valleys \citep{Pires1996}. This phenomenon appears in forecast optimizations beyond 5 days due to increasing gradient complexity with longer lead times \citep{heatwave_2024}. To overcome this problem, we gradually expand the optimization window size rather than fitting the entire trajectory at once. This proves to be a simpler gradient descent task and allows the algorithm to smoothly navigate what might be a complex loss manifold. The size of the optimization windows is arbitrary, but we choose an initial length of 2 days to allow forecast error to develop that is distinct from analysis error. Subsequent steps expand the window in 3-day increments, which works effectively based on empirical testing. \citet{Swanson-1998} implement a similar strategy for a four-dimensional variational data assimilation solver, also noting the performance improvement offered by progressively assimilating the total available data. They refer to this method as quasi-static, reflecting the stepwise adjustment of the assimilation window.

Details of the algorithm used to produce one set of optimized initial conditions (an ``optimal'') are as follows: 
\begin{enumerate} 
    \item Given a set of inputs, produce a forecast for the optimization window size. On the first pass, the initial window size is 2 days, and we initialize the forecast with the ERA5 analysis. Every subsequent epoch and window size starts with the optimal computed on the previous step. 
    \item Calculate the forecast loss function by verifying against ERA5 at every step during inference.  
    \item Calculate the gradient of the loss function with respect to the two input times ($t$ and $t-6\,\text{hr}$) using the JAX framework.
    \item Update the inputs using the Adam optimizer for gradient descent \citep{kingma2017}, applying the loss gradient as per Eq. (\ref{eqn:increment}). 
    \begin{enumerate}
        \item Repeat steps 1 -- 4 for a specified number of epochs, then proceed to step 5.
    \end{enumerate}
    \item Increase the optimization window size in step 1 by 3 days, or desired amount.
    \begin{enumerate}
        \item Repeat steps 1 -- 5 until the maximum optimization window size is reached. 
        \item  In our experiments, we use 100 epochs for lead times less than 10 days, 50 epochs for lead times between 10 and 20 days, and 25 epochs for lead times greater than or equal to 20 days.
    \end{enumerate}
\end{enumerate}

In this study, the maximum optimization window length is 32 days, limited by the 80 GB memory of the NVIDIA A100 GPU. With two input timesteps and 128 output timesteps, the number of free input parameters is approximately 1.6\% of the maximum target trajectory. As the forecast duration increases, so does the gradient size, necessitating a smaller model, greater GPU memory, or a strategy to further extend the window length. We suspect that optimization past 32 days would yield modest additional improvement. With respect to the optimizer hyperparameters, the default values ($\beta_1$ = 0.9, $\beta_2$ = 0.999) and a learning rate of \(10^{-3}\) are used, which have been shown to be effective \citep{heatwave_2024}. We progressively reduce the learning rate to handle the increasingly complex gradient descent for optimization windows longer than 14 days. Since each trajectory has a unique optimal initial condition, no batching is performed. For variables constrained to non-negative values, such as specific humidity and precipitation, optimization occasionally yields small negative perturbations; these are clipped to zero with negligible impact on the results. All findings reported in this paper represent the clipped initial conditions. With end-to-end double-precision floating-point operations, each optimization requires around 4 hours on an NVIDIA A100 GPU.

\subsection*{Loss Function}
As in \citet{heatwave_2024}, we adopt the scalar loss function used to train GraphCast, a weighted mean squared error (MSE) that quantifies the difference between predicted and target outputs, averaged over time, variables, and spatial locations. For a predicted state $\hat{x}$ and verification state $x$, the loss is expressed as: 

\begin{equation}
\label{eq2:loss}
\resizebox{0.7\columnwidth}{!}{%
  $\displaystyle
  \mathcal{L}
  =
  \underbrace{\frac{1}{T_{\mathrm{time}}}
     \sum_{\tau=1}^{T_{\mathrm{time}}}}_{\text{lead time}}
  \;
  \underbrace{\frac{1}{\lvert G_{1.0^\circ}\rvert}
     \sum_{i\in G_{1.0^\circ}}}_{\text{spatial location}}
  \;
  \underbrace{\sum_{j\in J}}_{\text{variable level}}
  s_j\,w_j\,a_i
  \bigl(\hat x_{i,j}^{t_0+\tau}-x_{i,j}^{t_0+\tau}\bigr)^{2}
  $%
}
\end{equation}

\noindent In this equation, $w$ represents the weight by pressure level, $a$ is the grid-cell area, and $s$ is a standardization parameter computed from time differences in the GraphCast training data. For more details on these parameters and the loss function, refer to Section 4.2 of the original GraphCast paper \citep{lam2023learning}.

\section{Results}\label{forecast_performance}

\subsection*{Forecast Performance}
Each initial condition is optimized by the GraphCast loss function (Eq. \ref{eq2:loss}) to reduce cumulative global forecast error over a 14-day window, yielding a set of 732 optimized forecasts computed using 32-bit floating-point arithmetic (hereafter, ``single-precision''). We restrict the optimization to 14 days due to diminishing returns stemming from loss of numerical precision, as longer windows require increasingly fine adjustments to the initial condition. To explore extended forecast horizons, we compute a 61-member subset---every sixth day of 2020---using 64-bit floating-point arithmetic (hereafter, ``double-precision''). GPU memory constraints limit double-precision optimization to 32 days, but the process could continue further with sufficient computing resources.

\begin{figure}[htbp]
  \centering
  \includegraphics[width=\linewidth]{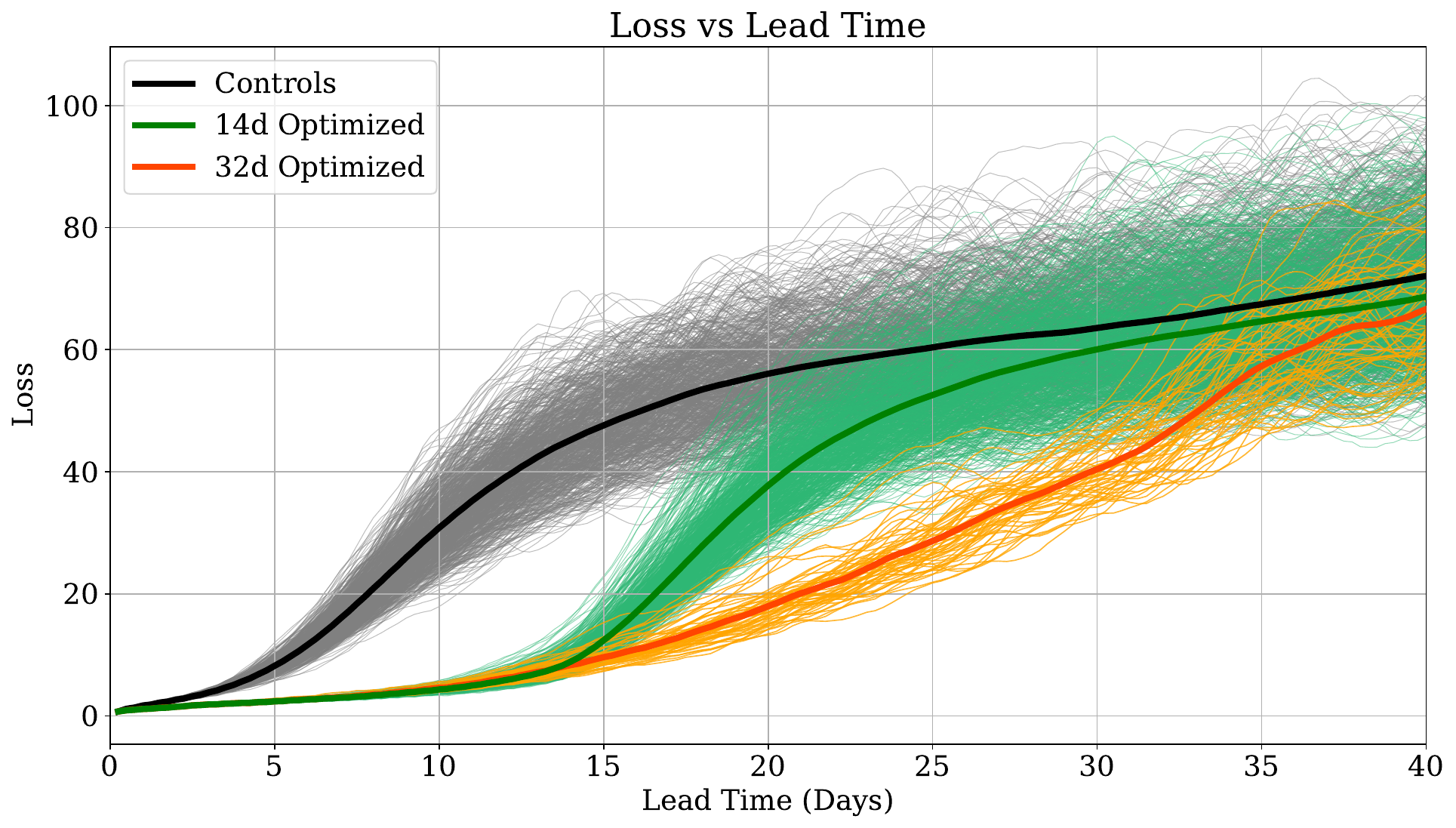}
  \caption{Weighted MSE as defined by Eq. (\ref{eq2:loss}) for all 732 control forecasts (black), 14-day optimized forecasts (green), and 32-day optimized forecasts (orange) during 2020.}
  \label{fig1:loss}
\end{figure}

When the loss is measured at ten days, the single-precision sample mean (green) shown in Fig.~\ref{fig1:loss} displays an 86\% reduction in weighted MSE compared to the control sample mean (black). Surprisingly, there are no failures. Each initialization time can be substantially optimized, with a minimum improvement of 77\% and a maximum of 91\%. Given that all 732 forecasts exhibit considerable error reduction up to 14 days, the technique appears effective for a wide range of atmospheric states during all seasons. This consistency is notable given the regime dependence of forecast skill implied by the ``predictability of predictability'' and operational forecasting bust studies \citep{dalcher1987, palmer1988, rodwell2013, Lillo2017, McLay2022}. Beyond the 14th day---the longest window for single-precision optimization---error growth returns to a rate that mirrors the control at earlier times until the two eventually merge near 30 days. 

The double-precision results (orange), optimized to 32 days, show a reduction in error relative to the control forecasts after the single-precision optimizations fail. The control and double-precision sample means have approximately equal error at 5 and 15 days, respectively. Errors grow at a nearly uniform exponential rate, with a doubling time of 5.8 days from day 2.5 until day 14 for both the single and double-precision sample means. This rate should not be interpreted as the intrinsic error doubling time of the atmosphere; rather, it reflects a combination of error growth from both the model and the initial conditions that remain resistant to the optimization procedure. Ultimately, the double-precision error growth rate gradually decreases as error saturates at the control value around day 37. Like the single-precision results, the double-precision curve also exhibits a subtle increase in error growth rate after the optimization window ends (day 32). The constant doubling time of errors is clearer when plotted with a logarithmic y-axis (Fig.~\ref{s2:30day_loss_log}), which also reveals an initial phase of elevated error growth (average doubling time of $\sim$1 day) between 6 and 24 hours, followed by a deceleration and a transition to the persistent 5.8-day doubling rate.

\begin{figure}[htbp]
  \centering
  \includegraphics[width=\linewidth]{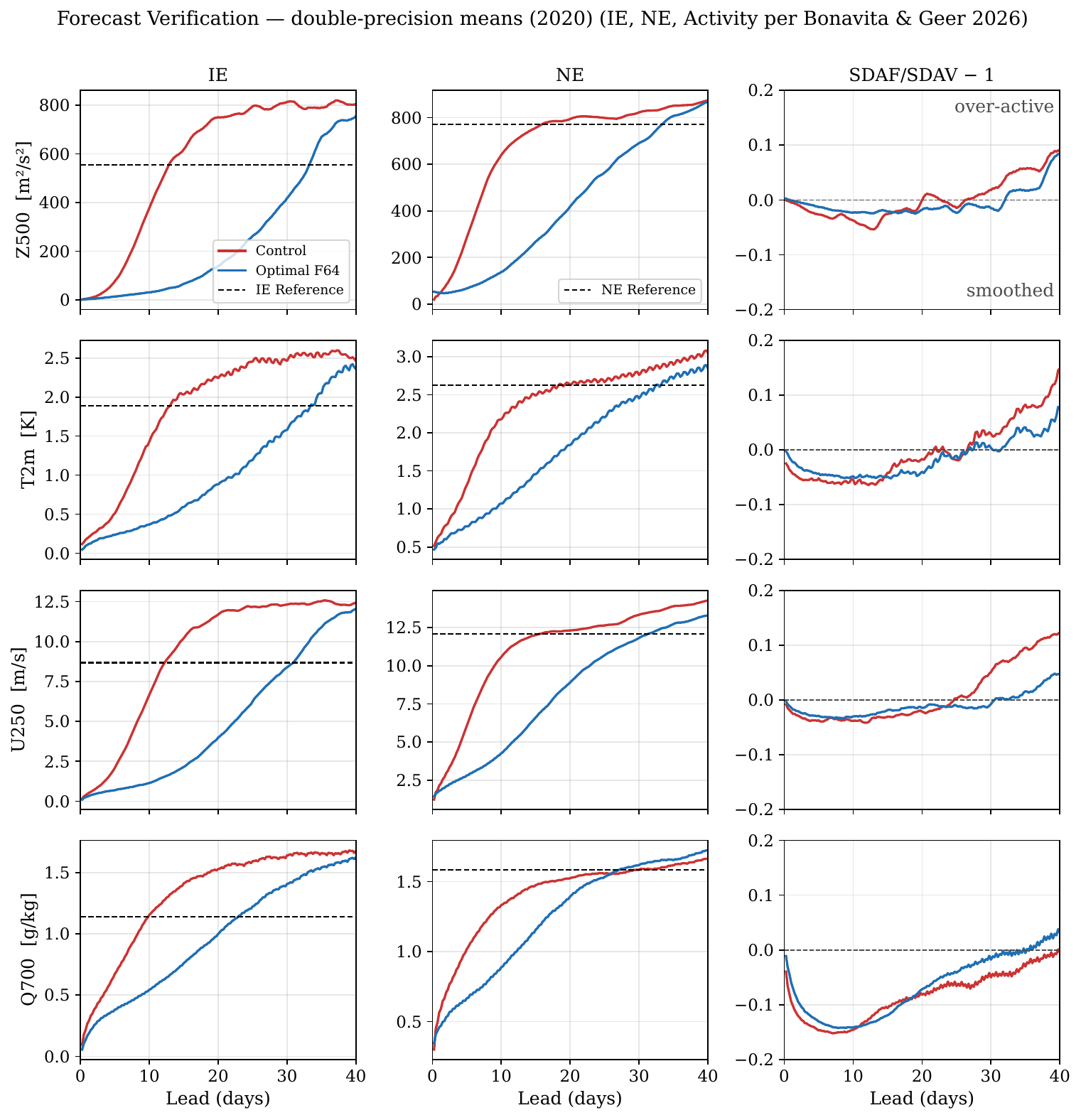}
  \caption{Forecast verification for unperturbed ERA5-control forecasts (red) and double-precision optimized forecasts (blue), averaged over 61 paired initialization dates during 2020. Rows show Z500, T2m, U250, and Q700; columns show information error (IE), noise error (NE), and activity bias $(\mathrm{SDAF}/\mathrm{SDAV})-1$ following \citet{ie_ne}. Dashed black lines show reference values based on the statistically significant ACC threshold described in App.~A2. SDAV and SDAF are the annual-mean anomaly standard deviations for ERA5 and the forecasts, respectively, referred to as \textit{activity}. Negative activity bias indicates under-active or smoothed forecasts; positive values indicate over-active forecasts.}
  \label{fig:ie_ne}
\end{figure}

It is worth noting that for exceptionally long forecasts---beyond 45 days---GraphCast is known to become unstable \citep{healpix2024}, and the results show early evidence of this in Fig.~\ref{fig1:loss}. The mean loss curves show a modest upward slope even beyond 35 days, never fully saturating. As a result, it is not clear from the loss exactly where forecast skill ends, so we compute the anomaly correlation coefficient (ACC) using the WeatherBenchX library \citep{rasp2023weatherbench,weatherbenchx2025} and find that for Z500 the anomaly correlation remains statistically different (see App.~A1) from the control at \( p \leq 0.01 \) to 33 days (Fig.~\ref{s3:acc}). Practical forecast skill, commonly defined as an ACC of 0.6 \citep{zhang2019predictability}, persists to 27.5 days. Geopotential is the best performing variable, consistent with enhanced sensitivity of ML-models to the height field \citep{bano_medina2025}. 

Since the optimization minimizes an MSE-based loss, one concern is that the extended skill could partly reflect smoothing or reduced forecast activity. To address this, Fig.~\ref{fig:ie_ne} presents information error (IE), noise error (NE), and forecast activity following \citet{ie_ne} (See App.~A2 for details). The optimized forecasts substantially reduce IE relative to the controls for 500mb geopotential height (Z500), 2-meter temperature (T2m), 250mb zonal wind (U250), and 700mb specific humidity (Q700), while forecast activity remains close to ERA5 for most variables. These results align with \citet{ie_ne}, who also find that GraphCast imparts the least amount of smoothing to the geopotential field. \citet{ie_ne} do not report on specific humidity---the most smoothed variable in our study---and only show these metrics to 15 days. Interestingly, GraphCast reacquires forecast activity beyond about 15 days in both the control and optimized forecasts, becoming consistently overactive after days 25--30. We suspect this is related to the known instability in the model at long lead times \citep{healpix2024}. Together, these results show that the optimized forecast improvements are not explained by smoothing, but instead are due to improved agreement with the verifying field. As an illustrative example, Fig.~\ref{fig_s8:z500_plots} displays the Z500 field for the worst-performing 30-day ERA5 optimized forecast and its corresponding control, depicting the evolution of synoptic-scale features in each forecast.

As a final sensitivity check, we repeat the full optimization procedure for a twelve-case subset using Modern-Era Retrospective analysis for Research and Applications Version 2 (MERRA-2) for both the initial conditions and verification targets \citep{merra2}. The MERRA-2 optimized forecasts show equal or slightly larger reductions in loss relative to their MERRA-2 controls, with error-growth behavior broadly comparable to the ERA5 optimizations (Fig.~\ref{fig:merra2_sensitivity}). These results on an independent reanalysis suggest that the forecast improvements seen with ERA5 are not a consequence of verifying against the reanalysis on which GraphCast was trained. 

\subsection*{Optimal Perturbation Sample-Mean Structure}

\begin{figure}[htbp]
  \centering
  \includegraphics[width=\linewidth]{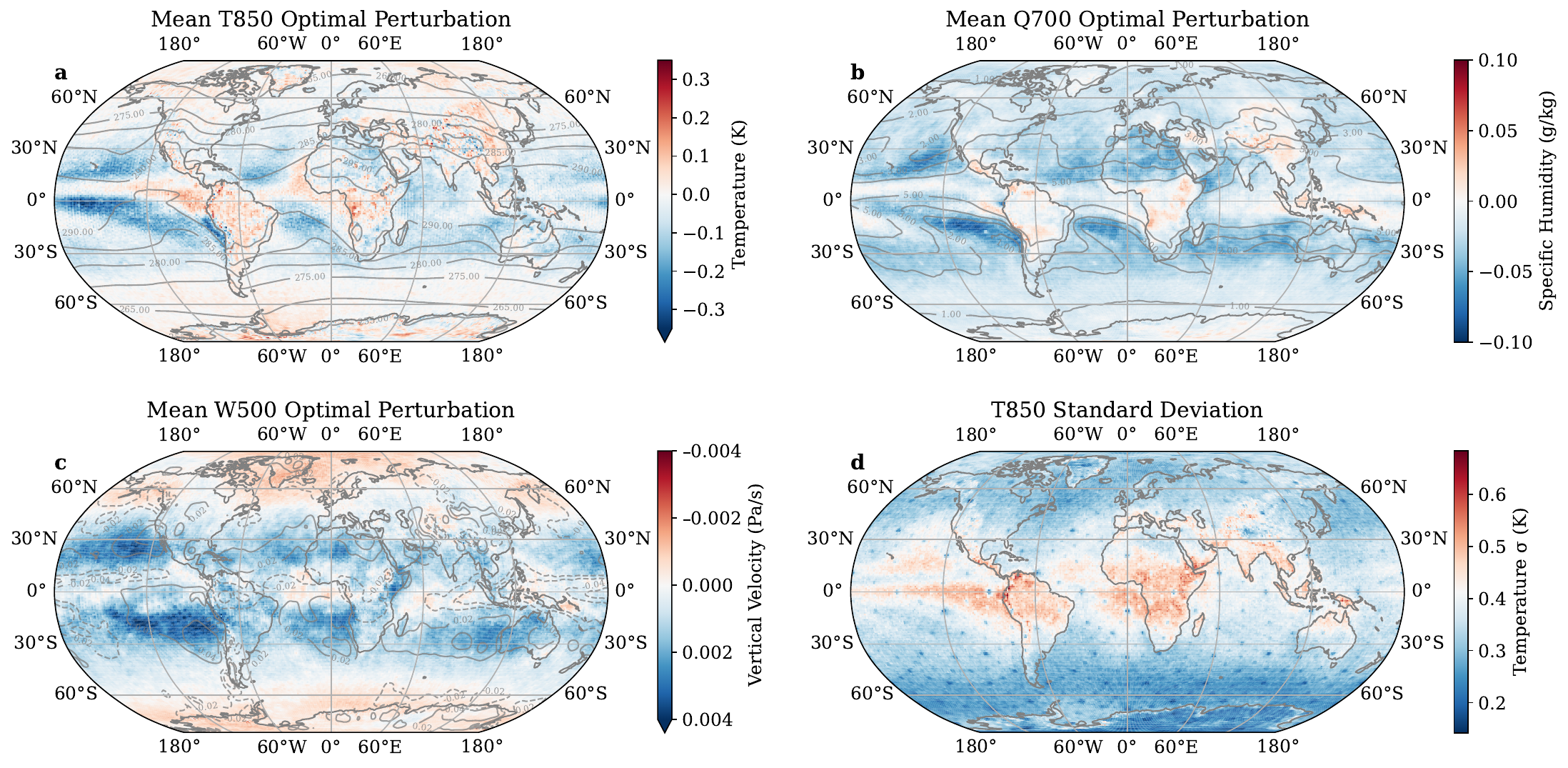}
  \caption{Sample-mean optimal perturbations averaged over 732 cases for (a) 850 hPa temperature, (b) 700 hPa specific humidity, (c) 500 hPa pressure vertical velocity (negative values indicate rising air); and (d) 850 hPa temperature sample standard deviation. Gray solid (dashed) contours represent the corresponding positive (negative) sample-mean values for ERA5.}
  \label{fig2:quad}
\end{figure}

The optimal perturbation sample-mean reveals coherent large-scale structure with greatest amplitude in the tropics and subtropics (Fig.~\ref{fig2:quad}). At 850 hPa, temperature perturbations (Fig.~\ref{fig2:quad}a) exhibit hemispherically symmetric cold anomalies over regions of subtropical stratocumulus cloud decks with warm anomalies along the Intertropical Convergence Zone (ITCZ), off the coast of Ecuador, and along the African west coast. These cold anomalies persist across seasons and shift meridionally, straddling the ITCZ (see Fig.~\ref{s5:seasonal_quadplot}). Cooling along the equatorial east Pacific may also be related to the 2020 La Niña event, possibly capturing the eastward extension of this event's cold tongue \citep{LA_NINA_2020}.

The 700 hPa specific humidity perturbations (Fig.~\ref{fig2:quad}b) correspond spatially to the temperature perturbations, showing drying of the mid-troposphere across subtropical oceans and moistening near the ITCZ, Central America, sub-Saharan Africa, and China. The central Indian Ocean and the Maritime Continent also exhibit increased moisture, consistent with the westward-shifted warm pool during La Niña \citep{LA_NINA_2020}. The mean summer moisture perturbations for East Asia and northern Australia suggest a strengthening of the monsoons in these locations (Fig.~\ref{s5:seasonal_quadplot}). Pressure vertical velocity perturbations at 500 hPa (Fig.~\ref{fig2:quad}c) further highlight the coherent structure of the GraphCast optimal initial conditions, with enhanced upward motion near the ITCZ, and increased subsidence throughout the subtropics consistent with mid-tropospheric drying. Increased upward motion relative to ERA5 characterizes the polar regions. 

To put the optimal perturbation amplitude range in perspective, Fig.~\ref{fig2:quad}d shows the 850 hPa temperature standard deviation. The most active regions generally mirror Fig.~\ref{fig2:quad}a, showing that perturbations along the ITCZ, northern South America, central Africa, and the Maritime Continent have the greatest mean magnitude relative to ERA5. This may suggest that these regions are under-resolved, or could reflect regional biases within GraphCast. Perturbation statistics reveal (see Table \ref{table1}) that the average magnitude of the perturbations is on the order of typical analysis error for all variables \citep[e.g.,][]{analysis1986,hakim2005,pena2014}. The icosahedral vertices and edges of the graph neural network appear as a subtle web of smaller standard deviation values, likely tied to GraphCast's encoding and decoding layers \citep{lam2023learning}. 

Overall, the sample-mean optimal structure represents a strengthening of the Hadley circulation, consistent with the weaker divergent wind component documented in ERA5 \citep{Li_2024} and radiosonde evidence that ERA5 underestimates upper-tropospheric poleward flow \citep{ziga2025}. This physically consistent structure of the sample-mean optimal is important because it suggests that GraphCast has learned relationships between variables and location, which are used to consistently correct multivariate errors. We also find that simply adding the sample-mean optimal perturbations to the control (ERA5) initial conditions reduces the loss by an average of 1-2\% over a 30-day period relative to the control forecasts (Fig.~\ref{s5:control_vs_optmean}). 

\begin{figure}[t]
  \centering
  \includegraphics[width=\linewidth]{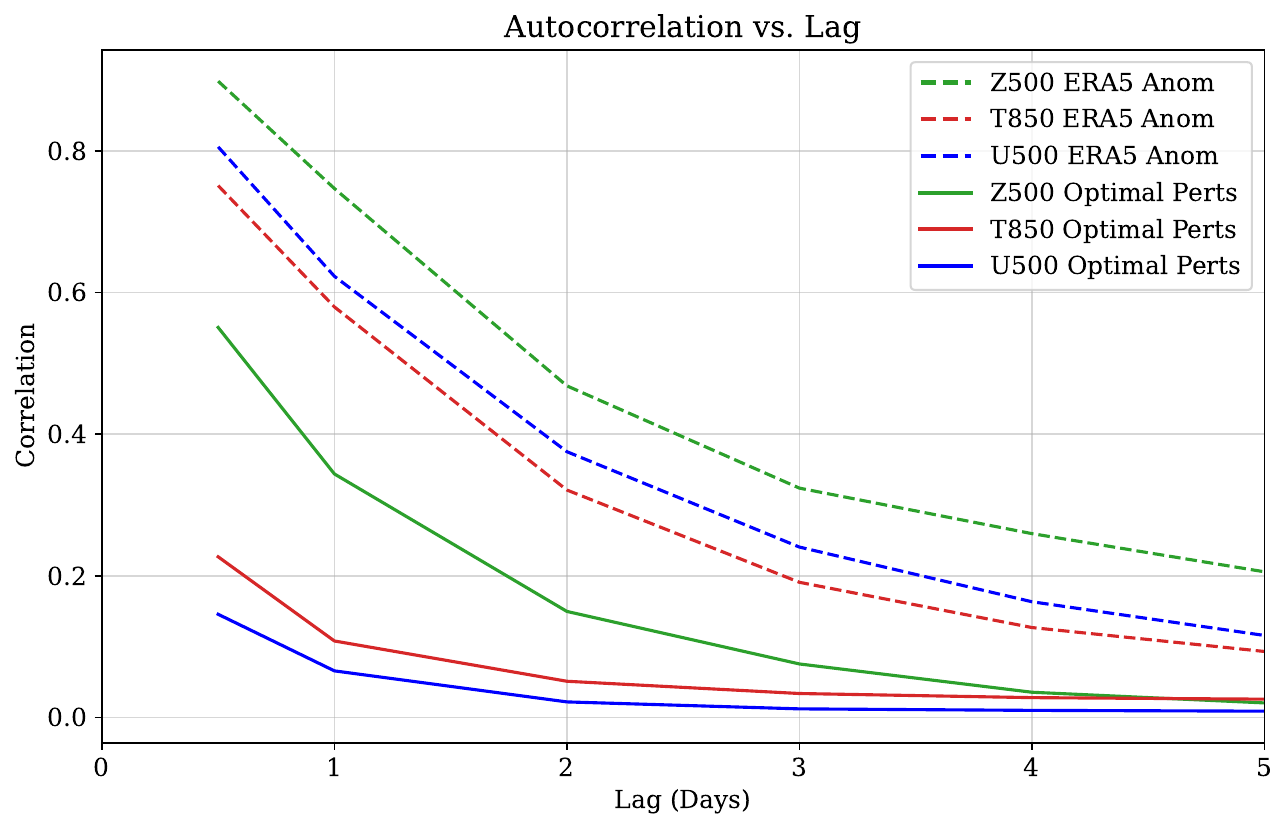}
    \caption{ERA5 2020 climatological anomaly and optimal perturbation autocorrelation as a function of lag for geopotential height (Z500), temperature (T850), and zonal wind (U500) at their respective pressure levels. Solid lines represent the global mean autocorrelation for the optimal perturbations while dashed lines show the ERA5 anomaly global-mean autocorrelation. The first correlation value is computed at 12 hours, consistent with the twice-daily optimization.}
  \label{fig3:autocorr}
\end{figure}

Analysis of the sample-mean perturbations in time and space reveals a distinct autocorrelation for each variable. In the global average, geopotential height has the most persistent autocorrelation for the optimal perturbations (Fig.~\ref{fig3:autocorr}). Temperature and zonal wind exhibit substantially lower initial autocorrelation values compared to geopotential height, but all three display similar e-folding times of 1.0 to 1.5 days. Zonal wind and specific humidity have enhanced autocorrelation in the tropics (not shown), whereas geopotential height has a more spatially uniform pattern. These results show that certain components of the optimized initial conditions---particularly the height field---have temporal persistence. However, the rapid decay in autocorrelation, especially for temperature and wind, indicates that a substantial portion of the adjustments are specific to the atmospheric state at the time of initialization. The ERA5 climatological anomaly mean autocorrelation shows significantly slower decay rates than those of the optimal perturbations (Fig.~\ref{fig3:autocorr}, dashed lines). Autocorrelation computation methodology can be found in App.~A3.

\subsection*{Cross-Model Forecast Validation} \label{pangu_section}

To assess how well the optimal initial conditions derived from GraphCast generalize to another model, we run forecasts with the 732 optimized initial conditions with the Pangu-Weather model \citep{bi2023}. Pangu-Weather is chosen for its distinctly different architecture, inference method, and spatial resolution. Unlike GraphCast, which uses two 6-hour time steps for autoregressive forecasts, Pangu-Weather employs different model weights trained for 1-, 3-, 6-, and 24-hour prediction intervals from a single time input. It also excludes vertical velocity and precipitation inputs and operates at a higher spatial resolution (0.25$^\circ$), necessitating interpolation of the optimized inputs (here, we use spherical harmonics, with zero padding at small scales). The model is designed to be used with the combination of time-interval weights that minimizes the number of inference steps required to reach a given forecast horizon; accordingly, we only use the 24-hour weights in this study.

\begin{figure}[t]
  \centering
  \includegraphics[width=\linewidth]{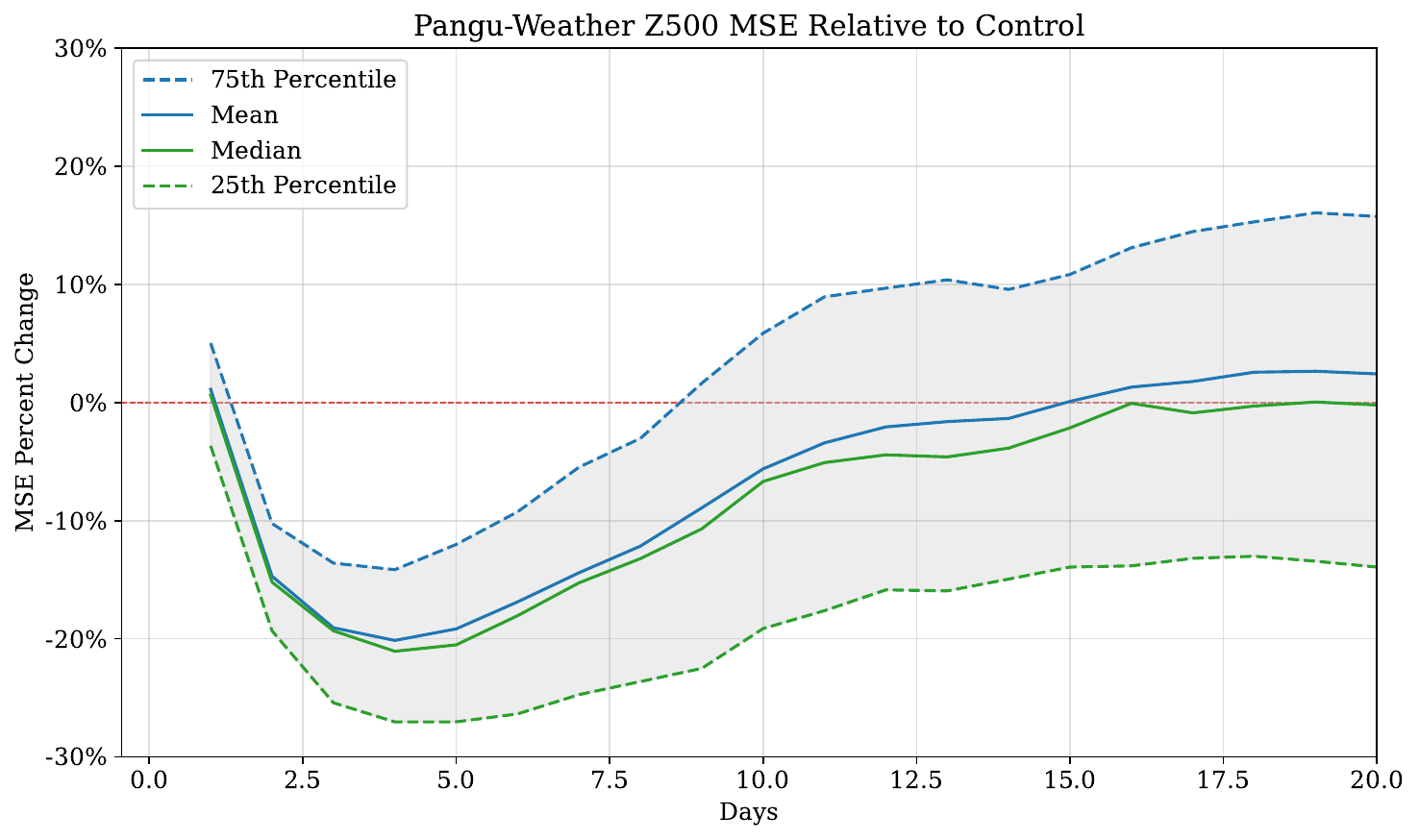}
    \caption{Mean, median, and inter-quartile range (25th--75th percentiles) of the relative change in MSE for Pangu-Weather forecasts using 732 GraphCast-optimized initial conditions. The control forecasts use 1.0° ERA5 data interpolated to a 0.25° grid for an equal comparison.}
  \label{fig4:pangu}
\end{figure}

Fig.~\ref{fig4:pangu} reveals an improvement for 500 hPa geopotential height throughout the 14-day optimization window, but greater variability (some forecasts are worse than the control) and much smaller improvement relative to GraphCast. When measured relative to the control forecast at 4 days, GraphCast optimal forecasts average a 62\% loss reduction, whereas Pangu-Weather forecasts average 21\%. This disparity further implies that the optimization procedure includes model-specific information, limiting the transferability of the resulting initial conditions across different forecasting systems. As a result, the true optimal analyses are unknown and will differ from those obtained here. The exact amount of model error relative to initial condition error is in general, state dependent, and a subject of future work.
 
Several factors likely contribute to smaller forecast improvements with Pangu-Weather. First, model-specific biases inherent to Pangu-Weather are likely different from GraphCast, diluting the impact of the optimized inputs. Second, GraphCast’s two 6-hour time-steps allow refinement of short-term tendencies, whereas Pangu-Weather---designed for single-step 24-hour forecasts---can only accept one of these optimized time levels, effectively receiving only half the information contained in the full optimization. Third, the absence of vertical velocity and precipitation in Pangu-Weather’s input also reduces the available optimized information. Finally, the interpolation process needed to project the optimized state onto the 0.25° grid introduces errors that may degrade the forecast performance given Pangu-Weather’s finer resolution.

\section{Discussion and Conclusions}
Using gradient-based optimization of initial conditions with the GraphCast model, we find weather forecast skill lasting roughly twice as long as the hypothesized limit of atmospheric predictability, exhibiting statistical significance in anomaly correlation to 33 days and useful skill up to 27.5 days. Cross-model validation with the Pangu-Weather model confirms that the optimized initial conditions yield significant but considerably smaller improvements, suggesting that the GraphCast-optimized initial conditions involve a blend of analysis improvement and model-specific error correction. Since GraphCast is trained on ERA5, separating model bias from reanalysis error remains ambiguous. The sample-mean optimal perturbations exhibit spatially coherent adjustments to the ERA5 reanalysis that broadly reflect an intensification of the Hadley circulation. 

We reiterate that traditional predictability studies have typically ascribed an intrinsic predictability limit by identifying the time at which two arbitrarily similar initial states become climatologically indistinguishable. In contrast, our study defines the limit for a single deterministic forecast as the time beyond which adjustments to the initial condition no longer reduce error. This distinction is critical, as prior studies of intrinsic atmospheric predictability either assume a perfect model and examine the divergence of nearby states, or adopt an ensemble approach, defining the predictability limit as the point at which ensemble spread approaches climatology. 

\begin{figure}[t]
    \centering
    \includegraphics[width=\linewidth]{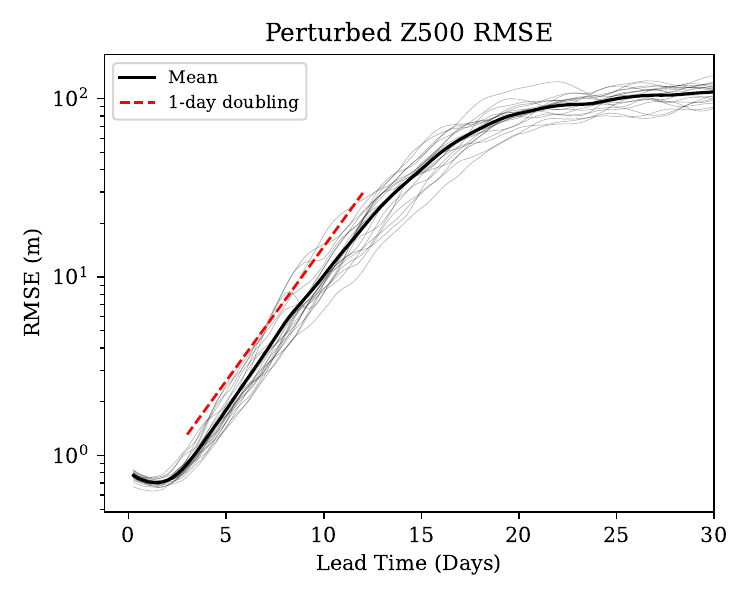}
        \caption{Perfect-twin experiment showing the Z500 RMSE over a 30-day forecast period for an ERA5 control and its perturbed twins. Perturbations are generated by adding white noise to the geopotential field, with mean amplitude set to approximately 25\% of the ECMWF operational analysis geopotential error \citep{pena2014}. The bold black curve denotes the ensemble mean; the red dashed curve serves as a 1-day doubling time reference; and gray curves represent the mean squared difference between each forecast pair--twenty total twins sampled approximately every 18 days during 2020. After initial decay, errors grow exponentially, approximating the leading Lyapunov vector in the model. Though lacking the initial rapid small-scale error amplification found in high-resolution traditional models, GraphCast is chaotic, as nearby states diverge exponentially with an error doubling time of $\sim$1 day from day 3 until saturation around twice the climatological value.}
      \label{fig5:gc_chaos}
\end{figure}

Previous perfect-twin experiments have shown that Pangu-Weather exhibits chaotic error growth comparable to physics-based models, with an error doubling time of approximately one day, but does not have the rapid small-scale error amplification typically associated with finite-time predictability \citep{selz2023}. In an equivalent perfect-twin experiment (Fig.~\ref{fig5:gc_chaos}), GraphCast behaves similarly, with initial decay followed by exponential growth. Within the prevailing view of atmospheric predictability, the absence of rapid initial error growth would typically be attributed to the lack of explicit small scales in this model. In that framework, rapid small-scale error growth is expected to limit atmospheric predictability to roughly two weeks and implies that no initial condition should produce skillful forecasts beyond that time \citep{palmer2014real}. An alternative view based on this study's findings is that large-scale error growth (e.g., Fig.~\ref{fig5:gc_chaos}) is controlled by large-scale initial-condition error. Our experiments show the existence of initial conditions that consistently yield skillful forecasts of the \textit{real} system---not a perfect-model twin---beyond 30 days. In our view, such results are most naturally explained if the atmosphere is at least this predictable at large scales, which suggests that rapid small-scale error growth may only weakly couple to larger scales. If this is true, the absence of strong small-scale amplification in ML models is not necessarily a deficiency, but an indication that atmospheric predictability may mostly be governed by large-scale dynamics.

While these findings underscore the existence of initial conditions with an extended intrinsic limit, they also raise the question of whether the improvements arise from the architecture of the ML model. In particular, one possible concern is that the optimization procedure may represent a form of adversarial attack on the neural network \citep[e.g.,][]{szegedy2014, deepfool, adv_space}. These attacks have been observed to change the output categorization of image classifier models through subtle single-pixel attacks \citep{adversarialexamples}. Adversarial examples are known to transfer across models under certain circumstances, meaning that testing results on different architectures (e.g., Pangu-Weather) may not always be protective. However, recent research has reframed these vulnerabilities, suggesting that what were once considered bugs are actually still predictive features of the model \citep{adv_feature,not_so_weird}. In any case, we hypothesize that several aspects of our method make it less prone to such attacks. 

First, in contrast to adversarial attacks, which aim to maximize output loss, our procedure seeks to minimize it. Second, across 732 distinct forecasts, a clear lower loss bound emerges: continued optimization---even using 10 times as many epochs---does not yield appreciable improvement. The uniformity of this behavior across numerous independent forecasts supports the robustness of the gradient descent procedure, which reliably converges to the lower loss bound despite significant variation across the full 2020 sample. Ultimately,  minimization of loss over such long trajectories through phase space (i.e., 56 consecutive states) demands highly precise and distinct optimal initial conditions. The emergence of physically interpretable, dynamically consistent large-scale structures in the optimal perturbations also argues against purely adversarial behavior.

Although GraphCast is deterministic, producing a single output for a given input, its predictions become blurred under multi-day loss minimization \citep{brenowitz2024practical,Charlton-Perez2024,bonavita2024}, damping small scale variability. As a result, mean squared error (MSE) measurements of machine learning models are not fully comparable to those of traditional deterministic physics-based models. That said, we note that GraphCast ranks among the least blurred ML weather models, with its 10-day forecast containing only 7\% less activity than the IFS analysis \citep{ecmwf_activity}. Moreover, anomaly correlation statistics, which mitigate overestimation of skill from blurring, also yield a limiting predictability timescale estimate of approximately 30 days. Future work could investigate models with higher resolution and with coupled ML atmosphere--ocean physics \citep[e.g.,][]{aurora, cresswell2025_earth}. Because the optimization technique can induce rapidly decaying noise in the optimal initial conditions, incorporating a regularization technique is another natural extension. It would also be interesting to evaluate the performance of ML-derived optimal initial conditions in traditional physics models. Collectively, our findings suggest the prospect of longer intrinsic atmospheric predictability. They do not yet provide a means of identifying optimal initial conditions in real-time, which would be required before realizing any operational forecasting gains.

\clearpage
\section*{Acknowledgments}
We acknowledge high-performance computing support from the Casper cluster (\url{https://doi.org/10.5065/qx9a-pg09}) provided by NCAR's Computational and Information Systems Laboratory, sponsored by the National Science Foundation. The Copernicus Climate Data Store provided access to ERA5. This research was supported by grants 2023-4715 from the Heising-Simons Foundation and 2501400 from the National Science Foundation. We thank Chris Snyder (NCAR), Matthew Chantry (ECMWF), Christian Lessig (ECMWF), Peter Dueben (ECMWF), Massimo Bonavita (ECMWF), and Dominik Stiller (UW) for conversations related to this work. Comments from three anonymous reviewers are gratefully acknowledged for improving the content and readability of the paper. Anthropic and OpenAI products were utilized to debug code, generate figures, and refine text.

%
%
\section*{Data Availability Statement}
All ERA5 initial condition data required to perform the optimizations in this study are available from the Copernicus Data Store (\url{https://doi.org/10.24381/cds.143582cf}) \citep{hersbach2017era5}. MERRA-2 data is available through the NCAR Research Data Archive \citep{merra2}. All optimal initial conditions produced by this study will be made available on Hugging Face Datasets at the time of publication. Code required to operate GraphCast and Pangu-Weather can be found at \url{https://github.com/google-deepmind/graphcast} \citep{lam2023learning} and \url{https://github.com/198808xc/Pangu-Weather} \citep{bi2023}, respectively. Initial condition optimization code is available at \url{https://github.com/tvonich/gc-initial-condition-optimization} \citep{heatwave_2024}.


\appendix
\renewcommand{\thesubsection}{\thesection\arabic{subsection}}  

\section{Additional Methodology}
\label{math_appendix}

\subsection{Effective Sample Size and Statistical Significance of ACC}
Forecast errors are temporally correlated, which motivates an estimate of the effective sample size for assessing statistical significance of long-lead forecasts. Therefore, we quantify the autocorrelation of Z500 forecast errors at 35-days, adjust the sample size accordingly using an effective sample size estimate, and compute corrected critical values for statistically significant anomaly correlation coefficients (ACC).

Let \( \{e_{n,L}\}_{n=1}^{N} \) be the sample of Z500 forecast errors  at a fixed lead time \(L\), where \(N=61\) is the number of double-precision forecasts and \(L=35\) days.  Define the lag-\(k\) autocorrelation at lead \(L\) by
\[
r_k(L) \;=\; \mathrm{corr}\bigl(e_{n,L},\,e_{n-k,L}\bigr).
\]
In particular, for a 6-day lag (\(k=1\)) when \(L=35\) days we observe
\[
r_1(35) \;=\; 0.08.
\]

Moreover, for lead times \(L=1,2,\dots,15\) days, the lag-1 autocorrelation remains below 0.01, indicating that forecast errors at these shorter leads are effectively independent.

To correct for this temporal correlation when testing anomaly correlation coefficients (ACC), we compute the effective sample size \citep{wilks2011}:
\[
N_{\rm eff}
\;=\;
\frac{N\,(1 - r_1)}{1 + r_1}
\;\approx\;
\frac{61\,(1 - 0.08)}{1 + 0.08}
\;\approx\;52.
\]
Using a one‐tailed \(t\)-test where $t_{\alpha,\nu}$ is the critical $t$-value corresponding to \(N_{\rm eff}-1 = 51\) degrees of freedom, the critical ACC values \(r_c\) satisfy
\[
r_c
\;=\;
\frac{t_{\alpha,\nu}}{\sqrt{t_{\alpha,\nu}^2 + \nu}},
\]
which yields
\[
r_c(p\le0.05)\approx0.23,
\qquad
r_c(p\le0.01)\approx0.32.
\]
Since our sample has no optimization failures, increasing \(N\) would likely raise \(N_{\rm eff}\), thereby marginally extending the maximum lead time for which ACC values are statistically significant.

\subsection{Information Error Benchmark and Activity Metrics}

Following \citet{ie_ne}, we compute information error (IE), noise error (NE), and forecast activity as follows.

Let $\mathbf{x}_f$, $\mathbf{x}_t$, and $\mathbf{x}_c$ denote the forecast, verifying analysis, and climatology fields, respectively, on a latitude--longitude grid. All IE and NE diagnostics in this study are computed globally. Let $m$ and $n$ index latitude and longitude, respectively, with latitude $\phi_m$. Area weights are proportional to grid-cell area, $w_m \propto \cos\phi_m$, and are normalized over the globe so that $\sum_{m,n} w_m=1$. We first form forecast and verifying climatological anomalies as
\begin{equation}
a_{f,m,n}=x_{f,m,n}-x_{c,m,n}, \qquad
a_{t,m,n}=x_{t,m,n}-x_{c,m,n}
\end{equation}
Their global area-weighted means are
\begin{equation}
\overline{a_f}=\sum_{m,n} w_m a_{f,m,n}, \qquad
\overline{a_t}=\sum_{m,n} w_m a_{t,m,n}
\end{equation}
The debiased forecast and debiased verifying anomalies are then
\begin{align}
d_{f,m,n} &= a_{f,m,n}-\overline{a_f} \\
d_{t,m,n} &= a_{t,m,n}-\overline{a_t}
\end{align}
This is the same anomaly-centering used in standard ACC verification, applied here over the global domain. The forecast and verifying anomaly activities are the area-weighted spatial standard deviations of the debiased anomaly fields,
\begin{equation}
\mathrm{SDAF}=\left(\sum_{m,n} w_m d_{f,m,n}^{\,2}\right)^{1/2}, \qquad
\mathrm{SDAV}=\left(\sum_{m,n} w_m d_{t,m,n}^{\,2}\right)^{1/2}
\end{equation}
where SDAF measures forecast activity and SDAV measures verifying-analysis activity. The anomaly correlation coefficient is
\begin{equation}
\mathrm{ACC}=
\frac{\sum_{m,n} w_m d_{f,m,n} d_{t,m,n}}
{\mathrm{SDAF}\,\cdot\mathrm{SDAV}}
\end{equation}

The forecast anomaly projected onto the verifying-anomaly direction has length $\mathrm{SDAF}\cdot\mathrm{ACC}$. The information error is the remaining error along that direction, and the noise error is the component orthogonal to it:
\begin{align}
\mathrm{IE} &= \left|\mathrm{SDAV}-\mathrm{SDAF}\cdot\mathrm{ACC}\right| \\
\mathrm{NE} &= \mathrm{SDAF}\left(1-\mathrm{ACC}^2\right)^{1/2}
\end{align}
Thus, IE measures the error in the forecast projection along the verifying-anomaly direction, penalizing insufficient or excessive aligned anomaly amplitude, while NE measures the component of the forecast anomaly orthogonal to the verifying anomaly. Together, these metrics separate improvements in anomaly information from reductions in forecast activity or noise, effects that can be conflated in RMSE and are not diagnosed by ACC alone because ACC is insensitive to anomaly amplitude.

For a climatology forecast, $\mathbf{x}_f=\mathbf{x}_c$, the forecast anomaly has zero amplitude and therefore zero projection onto the verifying anomaly. Thus, in the geometric limit,
\begin{equation}
\mathrm{IE}_{\mathrm{clim}}=\mathrm{SDAV}, \qquad \mathrm{NE}_{\mathrm{clim}}=0.
\end{equation}
Because $\mathrm{IE}<\mathrm{SDAV}$ is a weak no-information benchmark for a finite sample, we define IE and NE reference values using the ACC significance threshold from App.~A1. For simplicity, we evaluate this threshold assuming verification-matched activity, that is $\mathrm{SDAF}=\mathrm{SDAV}$, which is conservative for under-active forecasts and slightly permissive for over-active forecasts.
\begin{align}
\mathrm{IE}_{\mathrm{ref}} &= (1-\mathrm{ACC}_c)\,\mathrm{SDAV}=0.68\,\mathrm{SDAV}, \\
\mathrm{NE}_{\mathrm{ref}} &= \sqrt{1-\mathrm{ACC}_c^2}\,\mathrm{SDAV}\approx0.95\,\mathrm{SDAV}.
\end{align}
Here SDAV is the annual-mean ERA5 anomaly activity for each variable and level, computed from all 6-hourly 2020 verification times relative to the WeatherBench-2 1990--2019 climatology. Per-case scores are computed at fixed variable, level, and lead time, then averaged across cases.

\subsection{Autocorrelation Analysis}

Two populations of 12-hourly fields, each of length $N = 732$, are
compared in Fig.~\ref{fig3:autocorr}.

\paragraph{Optimal perturbations.}
For each initialization time $t_i$ ($i = 1, \ldots, N$) spanning all of
2020 at a 12~h interval, we define
\begin{equation}
  \delta \mathbf{x}(t_i) \;=\; \mathbf{x}^{\ast}(t_i) - \mathbf{x}_{\mathrm{ERA5}}(t_i),
  \label{eq:dx}
\end{equation}
where $\mathbf{x}^{\ast}(t_i)$ is the optimized initial condition
produced by the procedure of Section~2 and
$\mathbf{x}_{\mathrm{ERA5}}(t_i)$ is the corresponding ERA5 analysis.

\paragraph{ERA5 anomalies.}
The optimal perturbations target case-specific forecast error and resemble day-to-day atmospheric variability more than the seasonal cycle. At each initialization time $t_i$, the ERA5 climatological anomaly is defined as:
\begin{equation}
\mathbf{a}(t_i) \;=\; \mathbf{x}_{\mathrm{ERA5}}(t_i)
\;-\; \overline{\mathbf{x}}_{\mathrm{clim}}\!\left(d_i, h_i\right),
\label{eq:anom}
\end{equation}
where $\overline{\mathbf{x}}_{\mathrm{clim}}(d, h)$ is the WeatherBench2 ERA5 hourly climatology over 1990--2019, indexed by day-of-year $d$ and hour-of-day $h$ \citep{rasp2023weatherbench}. The quantities $d_i$ and $h_i$ denote the corresponding values at $t_i$. The 6-hourly WeatherBench2 ERA5 analyses are sub-sampled to 12~h to match the cadence of the optimal perturbations. The analyses and climatology both reside on the WeatherBench2 $1.5^{\circ}$ equiangular grid ($121 \times 240$). The optimal perturbations are conservatively regridded from their native $1^{\circ}$ GraphCast grid onto this grid so that both populations are compared at matched resolution.

\paragraph{Pointwise autocorrelation.}
At each grid point $(m,n)$ the lag-$k$ temporal
autocorrelation of a generic time series $y(t_i)$ is defined as 
\begin{equation}
  \rho_k(m,n)
  \;=\;
  \frac{\displaystyle \sum_{i=1}^{N-k}
        \bigl[y(t_i) - \mu\bigr]\bigl[y(t_{i+k}) - \mu\bigr]}
       {\displaystyle \sum_{i=1}^{N}
        \bigl[y(t_i) - \mu\bigr]^{2}},
  \quad
  \mu = \frac{1}{N}\sum_{i=1}^{N} y(t_i).
  \label{eq:rho}
\end{equation}
The lag $k$ is an integer number of 12~h samples, so lag in days is
$k/2$. Fig.~\ref{fig3:autocorr} uses
$k \in \{1,2,4,6,8,10\}$, corresponding to
$\{0.5,1,2,3,4,5\}$~days.

\paragraph{Spatial average.}
The curves in Fig.~\ref{fig3:autocorr} show a global spatial mean of
$\rho_k(m,n)$ weighted by $\cos\phi_m$ to correct for unequal grid area,
\begin{equation}
  \bar{\rho}_k
  \;=\;
  \frac{\displaystyle \sum_{m,n}
        \rho_k(m,n)\,\cos\phi_m}
       {\displaystyle \sum_{m,n} \cos\phi_m}.
  \label{eq:rho_bar}
\end{equation}

\clearpage

\section{Additional Figures and Tables}
\label{appendix_figs}



\begin{figure}[htbp]
\begin{center}
  \noindent\includegraphics[width=\textwidth]{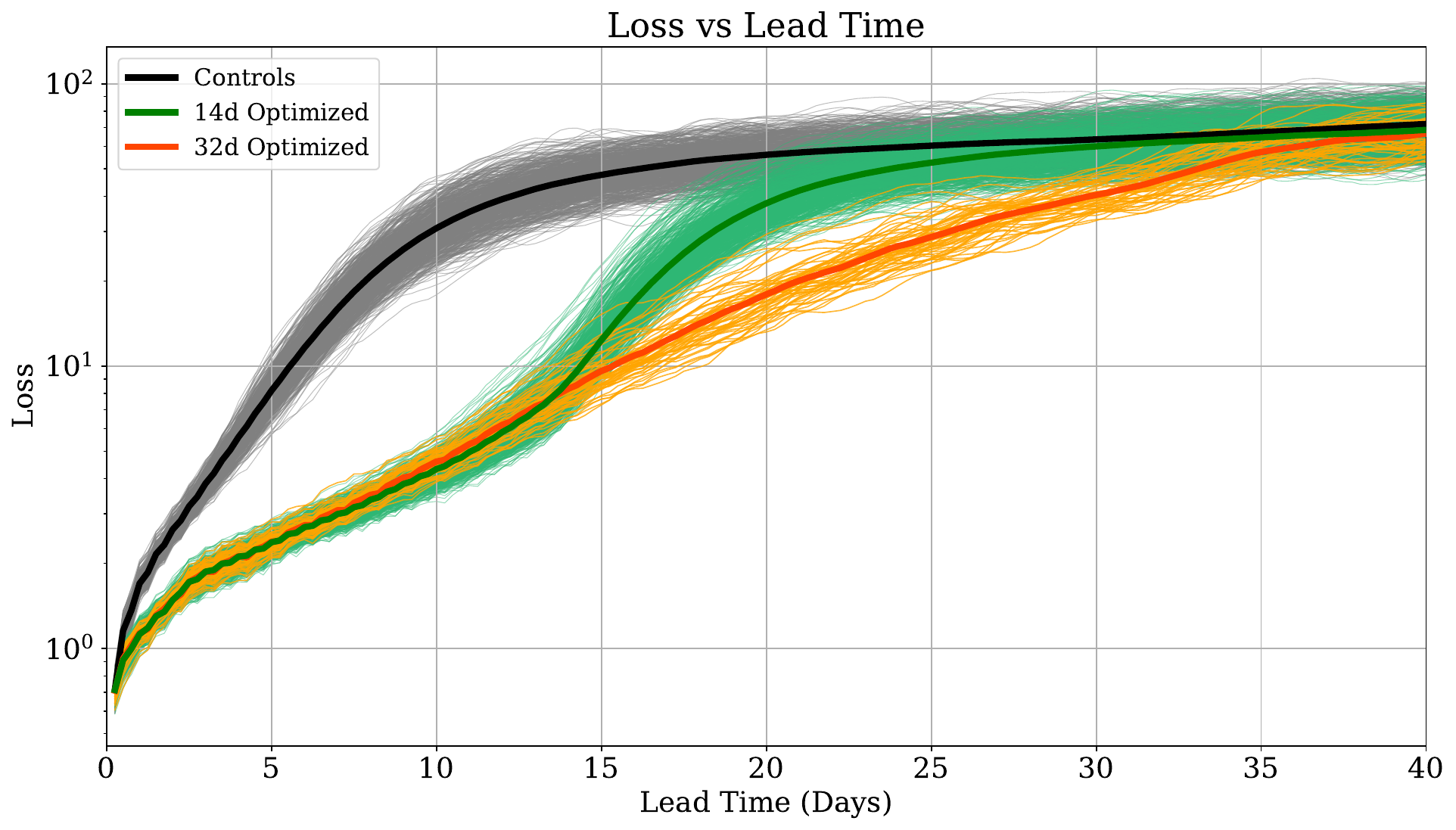}
\end{center}
  \renewcommand{\thefigure}{B1} 
  \caption{As in Fig. 1, but with a log y-axis. Error growth is initially fast but decelerates between 0 and 2.5 days, followed by a steady exponential growth rate thereafter. Recall that the green curves have been optimized to 14-days and return an error growth rate similar to the control beyond the optimization window length.} \label{s2:30day_loss_log}
\end{figure}
\clearpage

\begin{figure}[t]
\begin{center}
  \noindent\includegraphics[width=\textwidth]{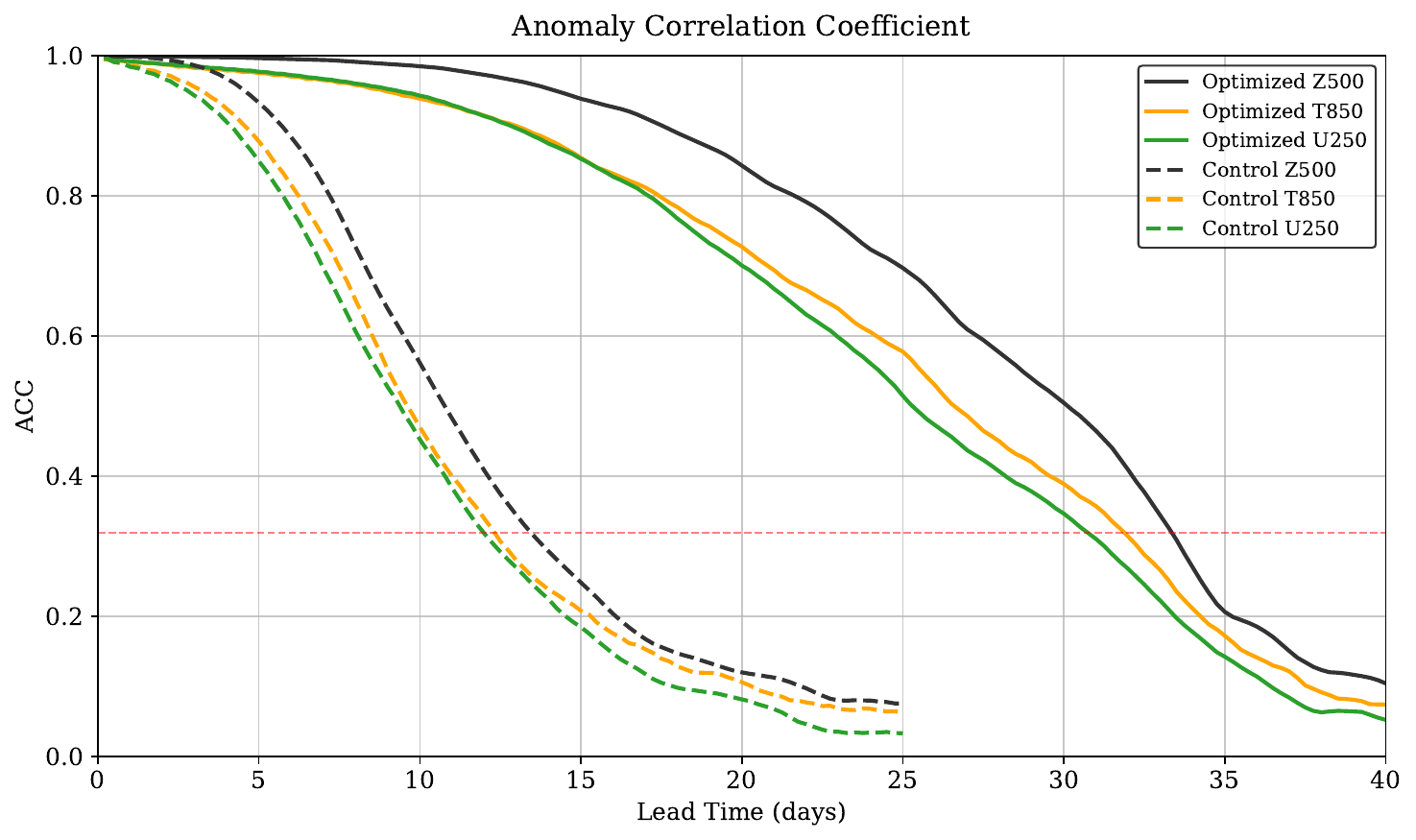}
\end{center}
  \renewcommand{\thefigure}{B2} 
  \caption{2020 global-mean anomaly correlation coefficient (ACC) for key variables and pressure levels. Solid lines represent optimal forecasts for 500 hPa geopotential height (black), 850 hPa temperature (orange), and 250 hPa zonal wind (green). Dashed lines show results for the corresponding control forecasts for the same variables. The red horizontal dashed line indicates the 0.32 threshold, above which ACC values are statistically significant ($p \leq 0.01$).} \label{s3:acc}
\end{figure}
\clearpage


\begin{figure}[t]
\begin{center}
  \noindent\includegraphics[width=\textwidth]{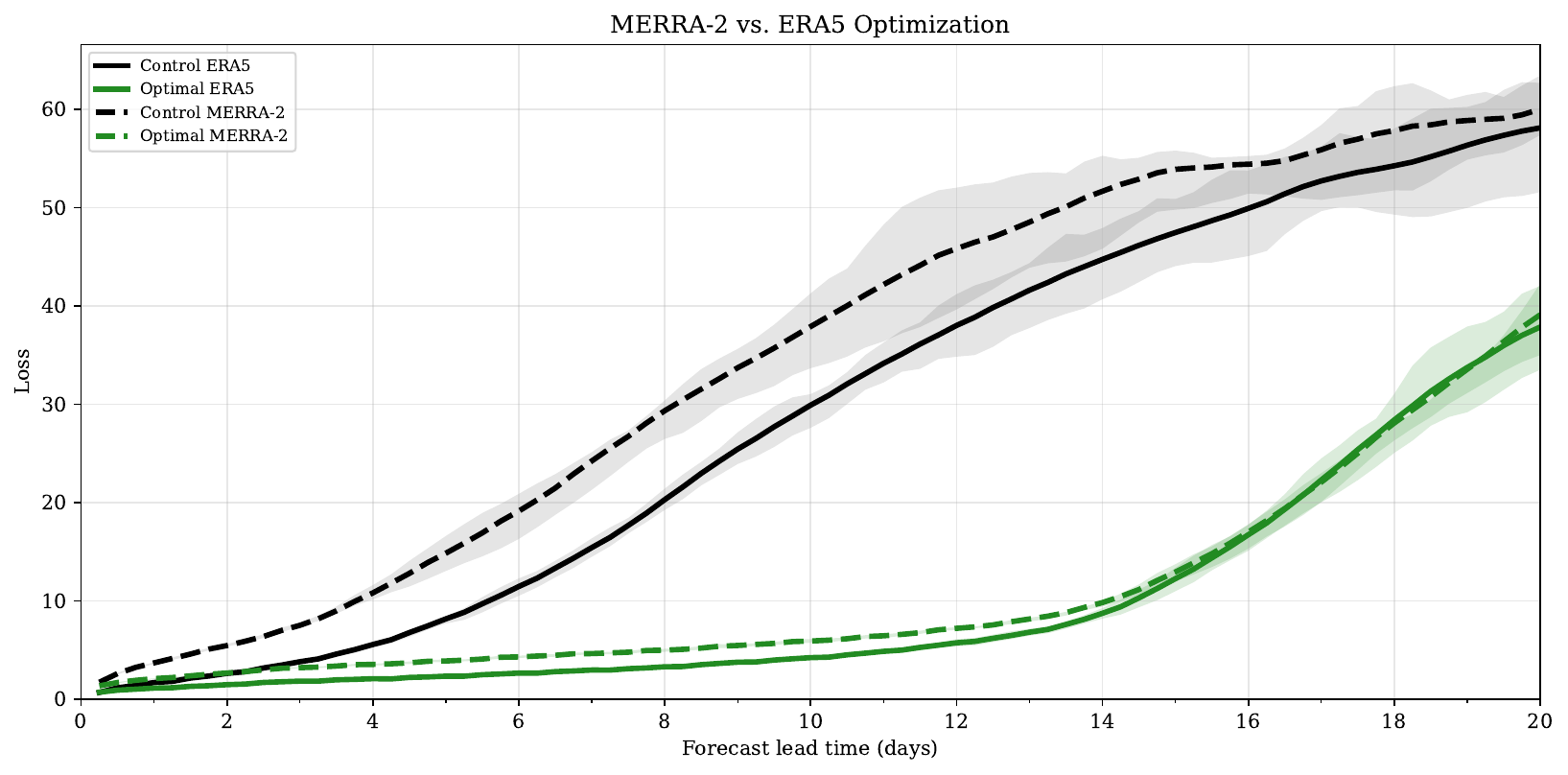}
\end{center}
  \renewcommand{\thefigure}{B3} 
\caption{Weighted mean squared error (GraphCast loss) for ERA5 and MERRA-2 control and optimized forecasts. Solid lines show ERA5 results (as in Fig.~\ref{fig1:loss}), and dashed lines show a twelve-case MERRA-2 optimization experiment using forecasts initialized on the first day of each month in 2020, with both the initial conditions and verification targets taken from MERRA-2. Thick lines show sample means, and shaded envelopes show the 25th--75th percentile range across initialization times at each forecast lead. The MERRA-2 optimized forecasts show large loss reductions relative to their controls, with error growth broadly comparable to the ERA5 optimized forecasts. The larger MERRA-2 errors likely reflect a combination of regridding differences, incomplete data below terrain elevation in MERRA-2 (infilled with ERA5), real analysis differences at initialization, and GraphCast's ERA5-specific biases.}
\label{fig:merra2_sensitivity}
\end{figure}

\begin{figure}[t]
\begin{center}
  \noindent\includegraphics[width=\textwidth]{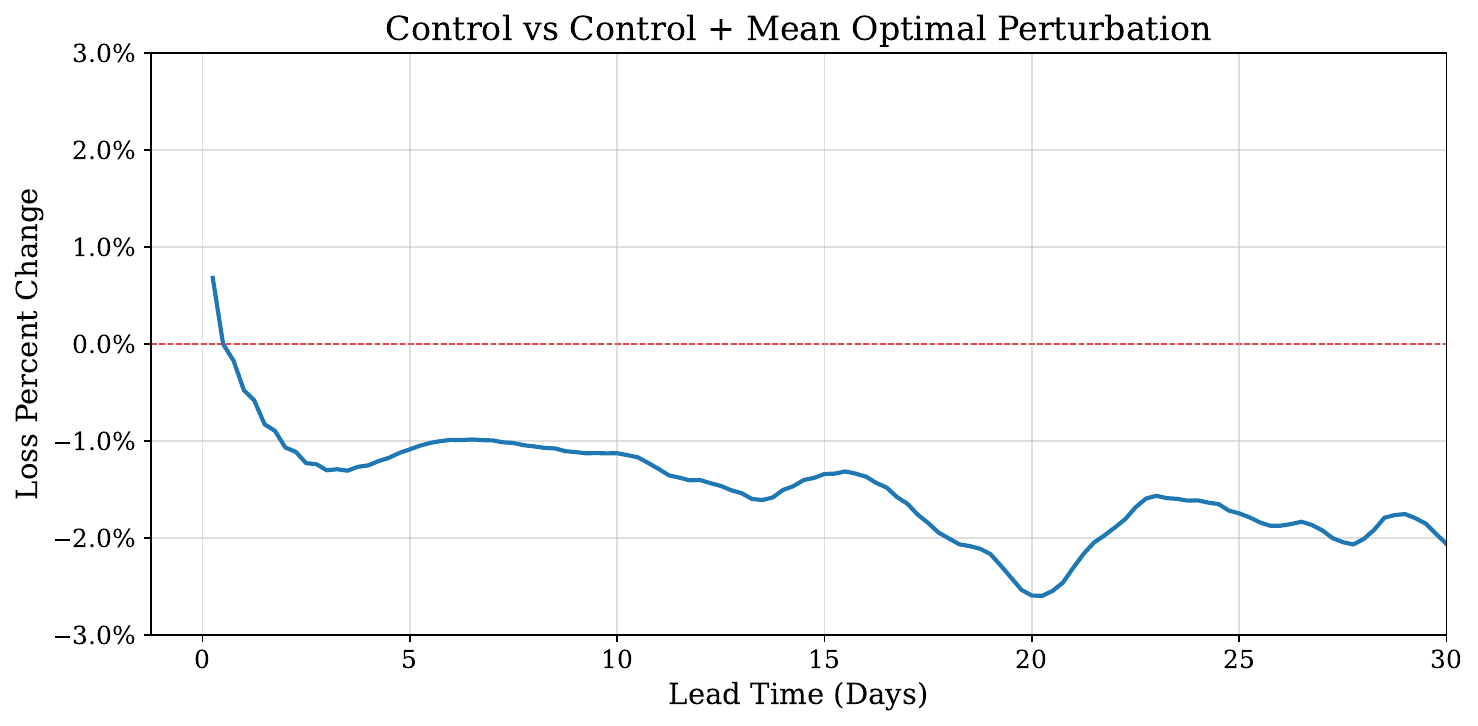}
\end{center}
  \renewcommand{\thefigure}{B4} 
    \caption{Results for forecasting experiments where the sample-mean optimal perturbation (as illustrated in Fig. 2) is added to the 732 unperturbed initial conditions from ERA5. The forecast loss is then computed, averaged, and compared to the original control loss seen in Fig. 1. The result is a $\sim$1--2\% average improvement across the 30-day forecast window.}
  \label{s5:control_vs_optmean}
\end{figure}

\begin{figure}[t]
\begin{center}
  \noindent\includegraphics[width=\textwidth]{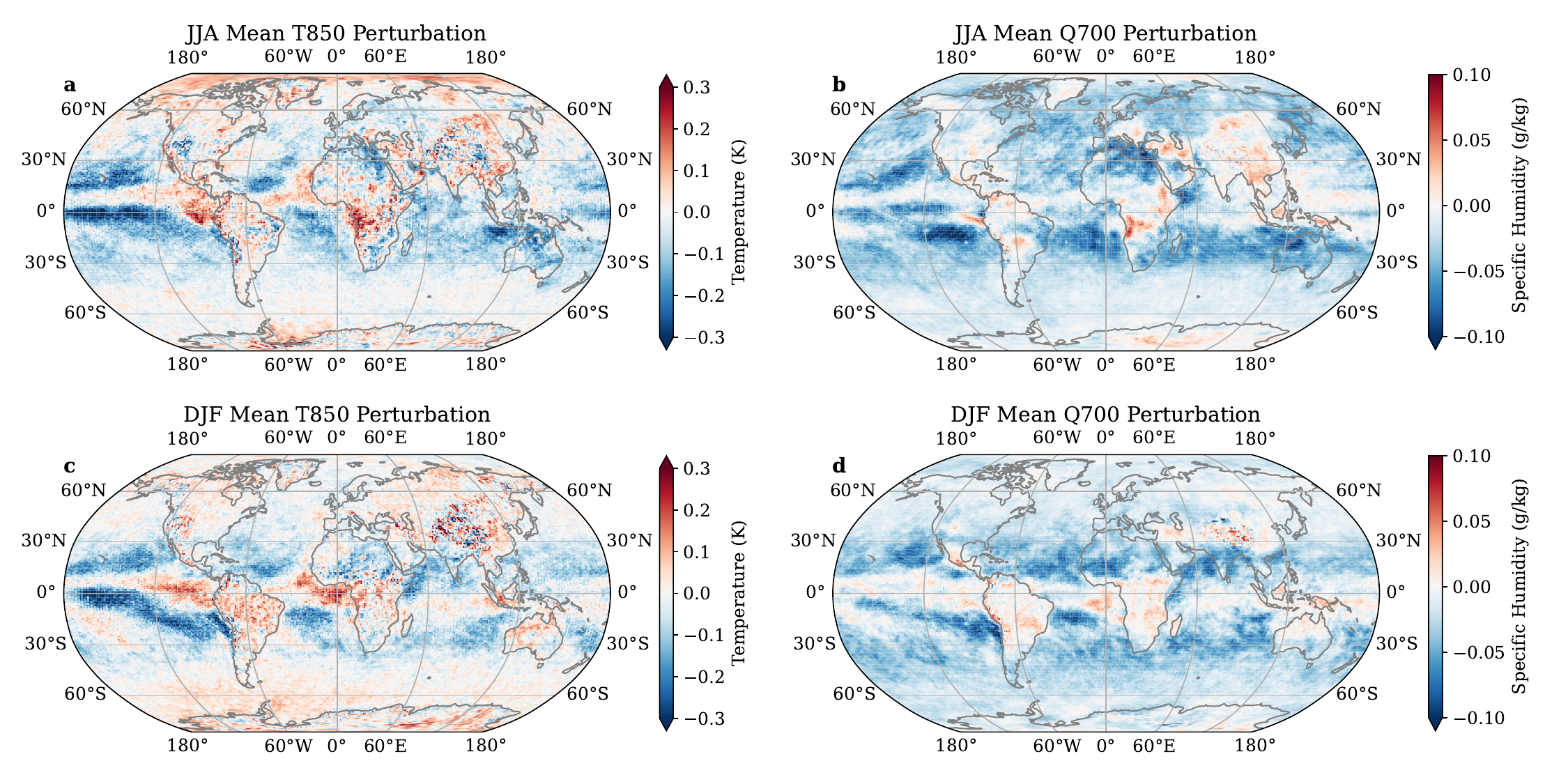}
\end{center}
  \renewcommand{\thefigure}{B5}
    \caption{Seasonal-mean optimal perturbations for boreal summer (JJA, top row) and boreal winter (DJF, bottom row), for T850 (left column) and Q700 (right column). The dominant features of the annual mean (Fig.~\ref{fig2:quad}) persist in both seasons: tropical Pacific T850 cooling and broad Q700 drying across the tropics and subtropics. Seasonal differences are qualitatively and physically consistent: Northern Hemisphere continental T850 warm anomalies over North America, Europe, and northern Asia are markedly stronger in JJA, while Australia and southern South America cool in JJA (austral winter); Q700 moistening over East Asia in JJA and northern Australia in DJF corresponds to the active phases of the East Asian and Australian summer monsoons, respectively; and the subtropical subsidence drying has a southern bias during DJF and northern bias during JJA, suggesting a year-round underestimation of the Hadley circulation in ERA5.}
  \label{s5:seasonal_quadplot}
\end{figure}

\begin{figure}[t]
\centering

\newcommand{\panelH}{0.105\textheight} 
\newcommand{\panel}[2]{%
  \includegraphics[width=\textwidth,height=\panelH,keepaspectratio]{#1}\par
  {\scriptsize #2}\par\vspace{0.35em}%
}

\begin{minipage}[t]{0.32\textwidth}\centering
  \panel{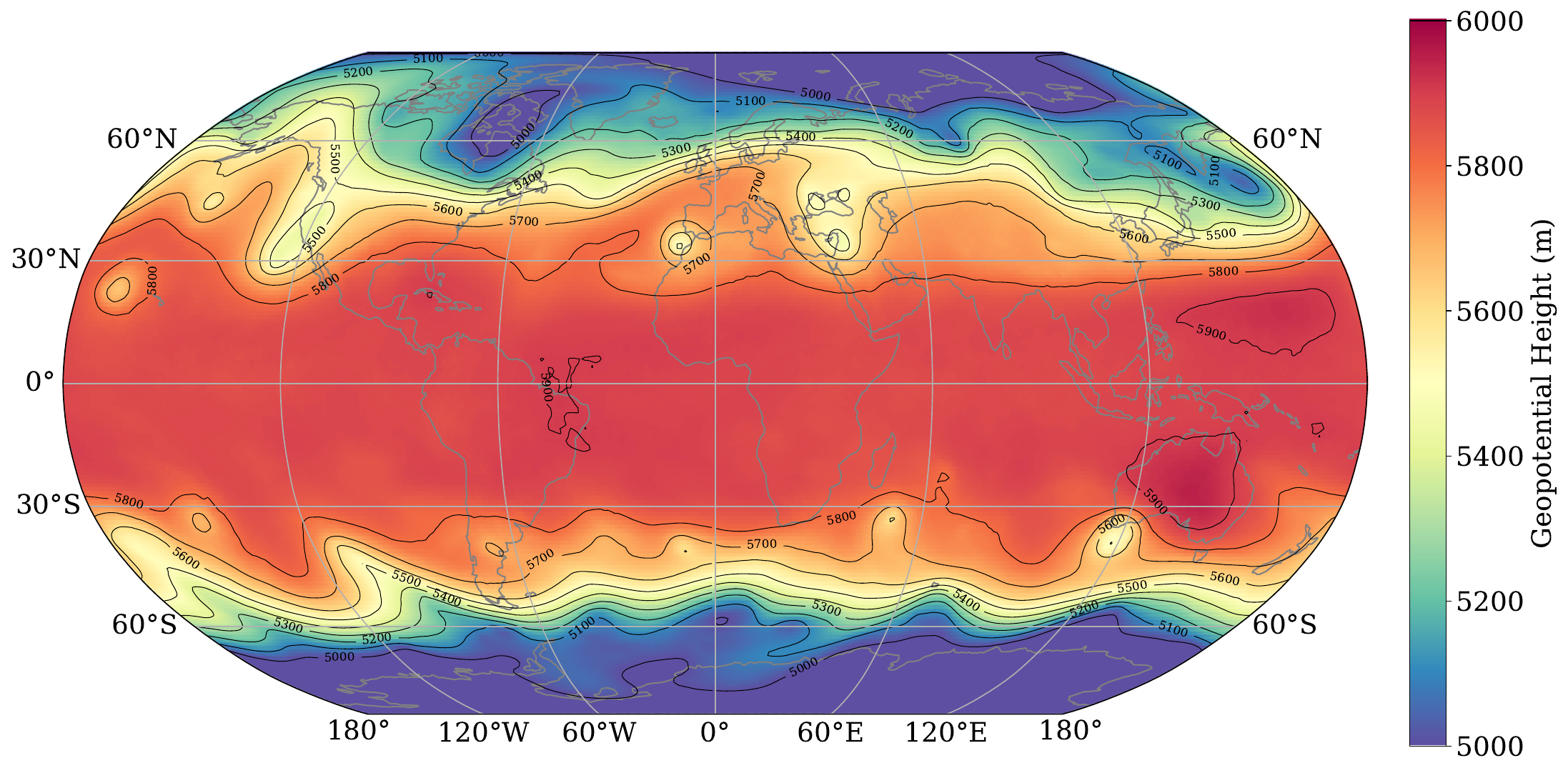}{ERA5 00Z, March 18, 2020}
  \panel{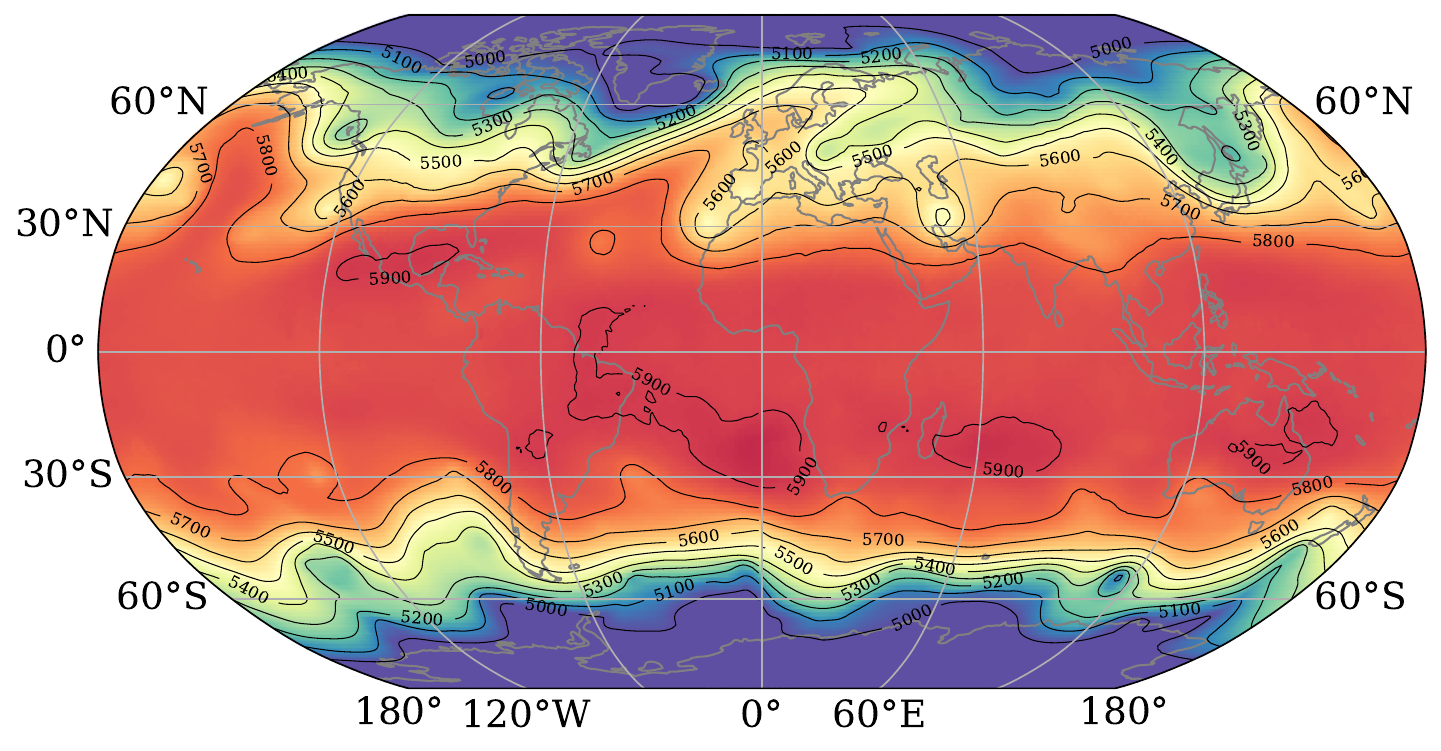}{ERA5 00Z, March 23, 2020}
  \panel{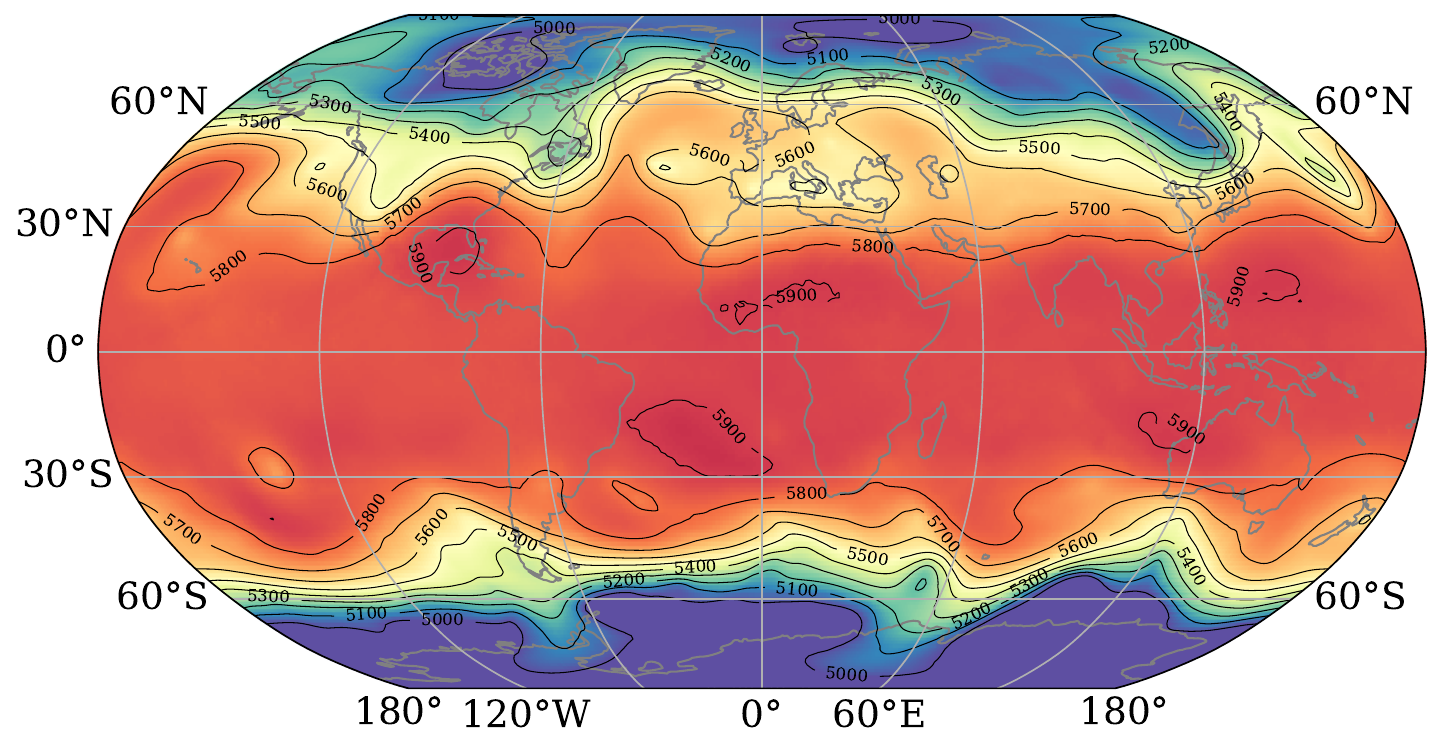}{ERA5 00Z, March 28, 2020}
  \panel{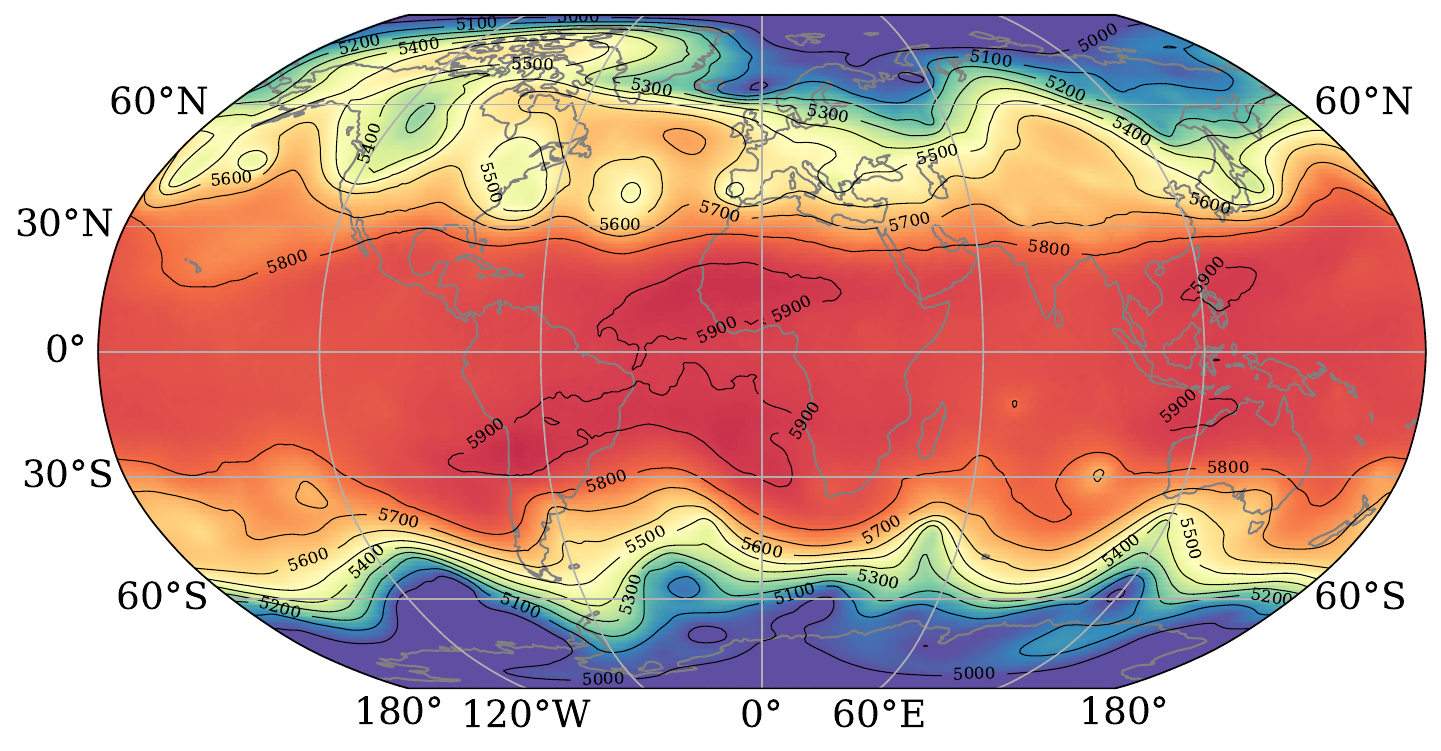}{ERA5 00Z, April 2, 2020}
  \panel{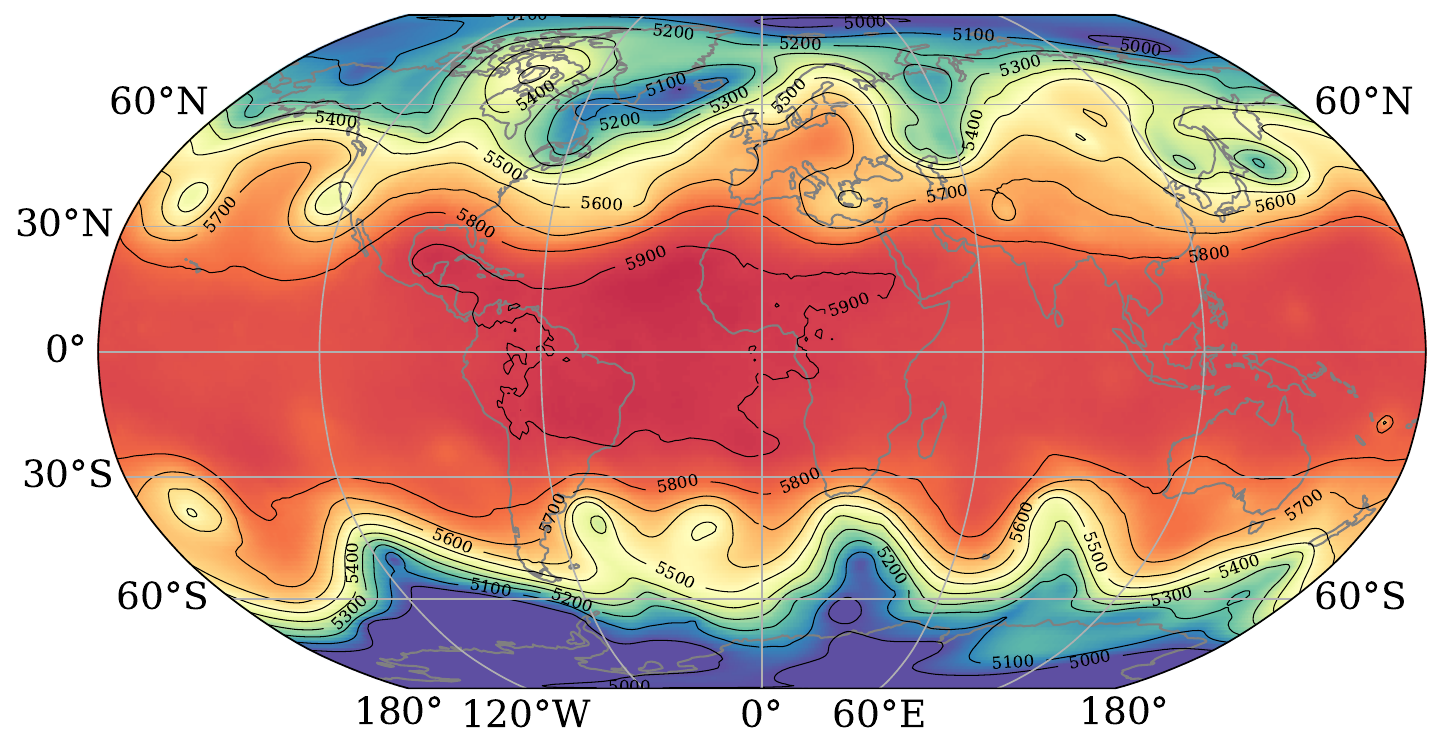}{ERA5 00Z, April 7, 2020}
  \includegraphics[width=\linewidth,height=\panelH,keepaspectratio]{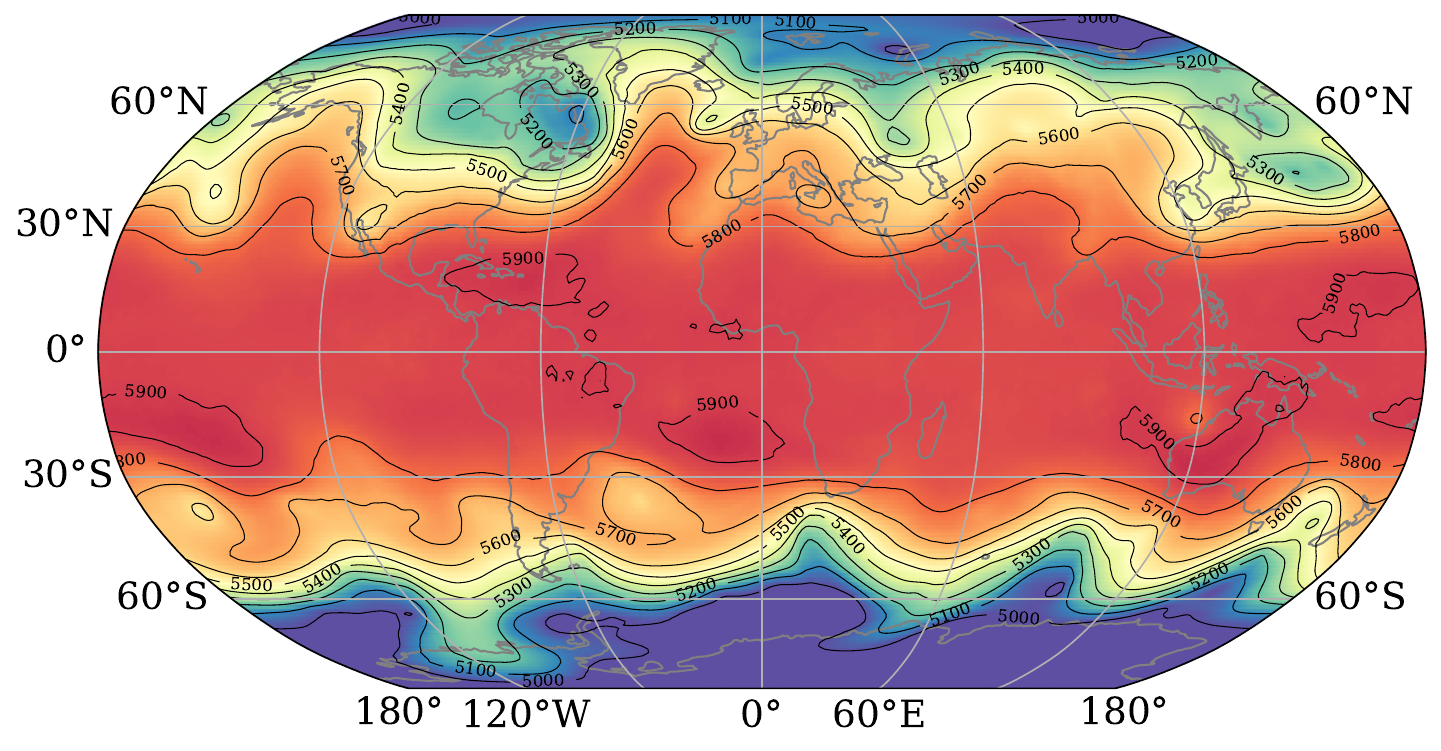}\par
  {\scriptsize ERA5 00Z, April 12, 2020}
\end{minipage}\hfill
\begin{minipage}[t]{0.32\textwidth}\centering
  \panel{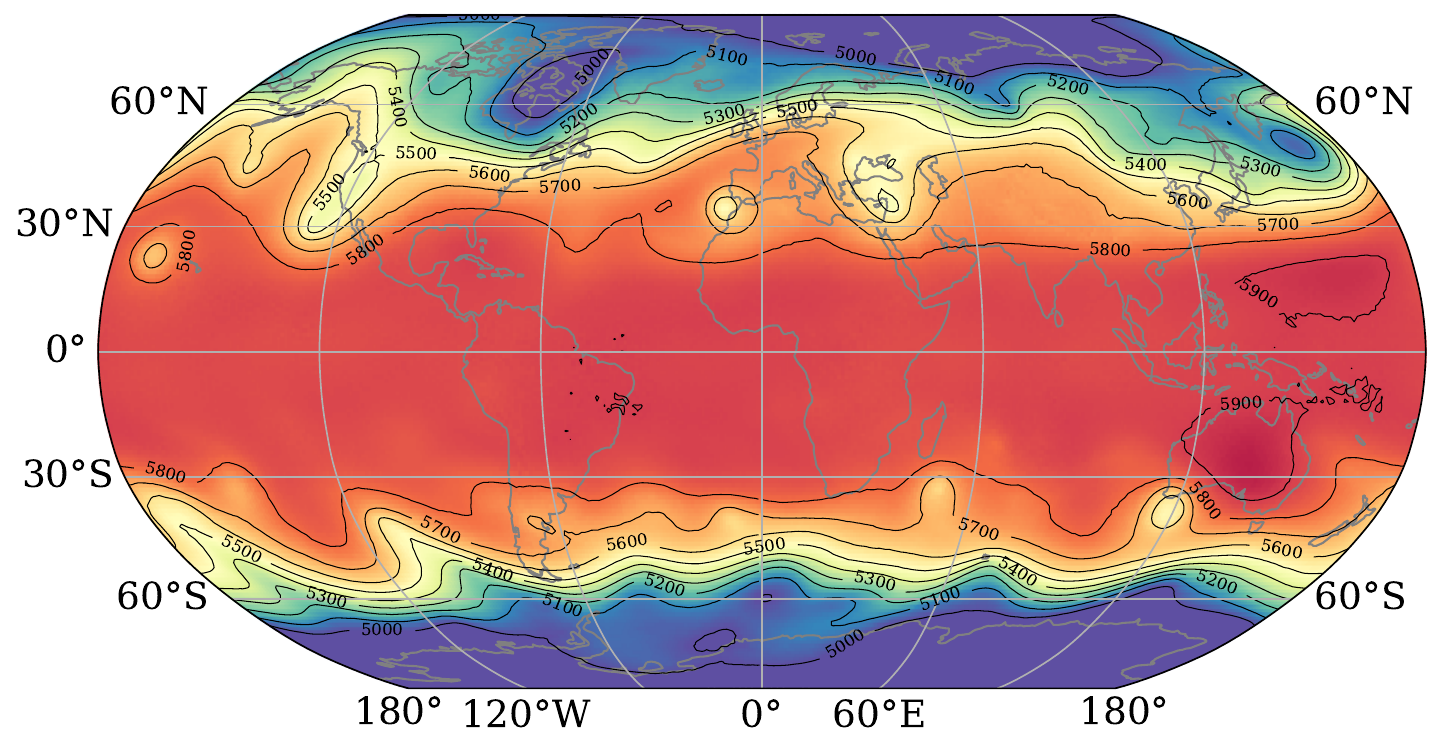}{5-day optimized forecast}
  \panel{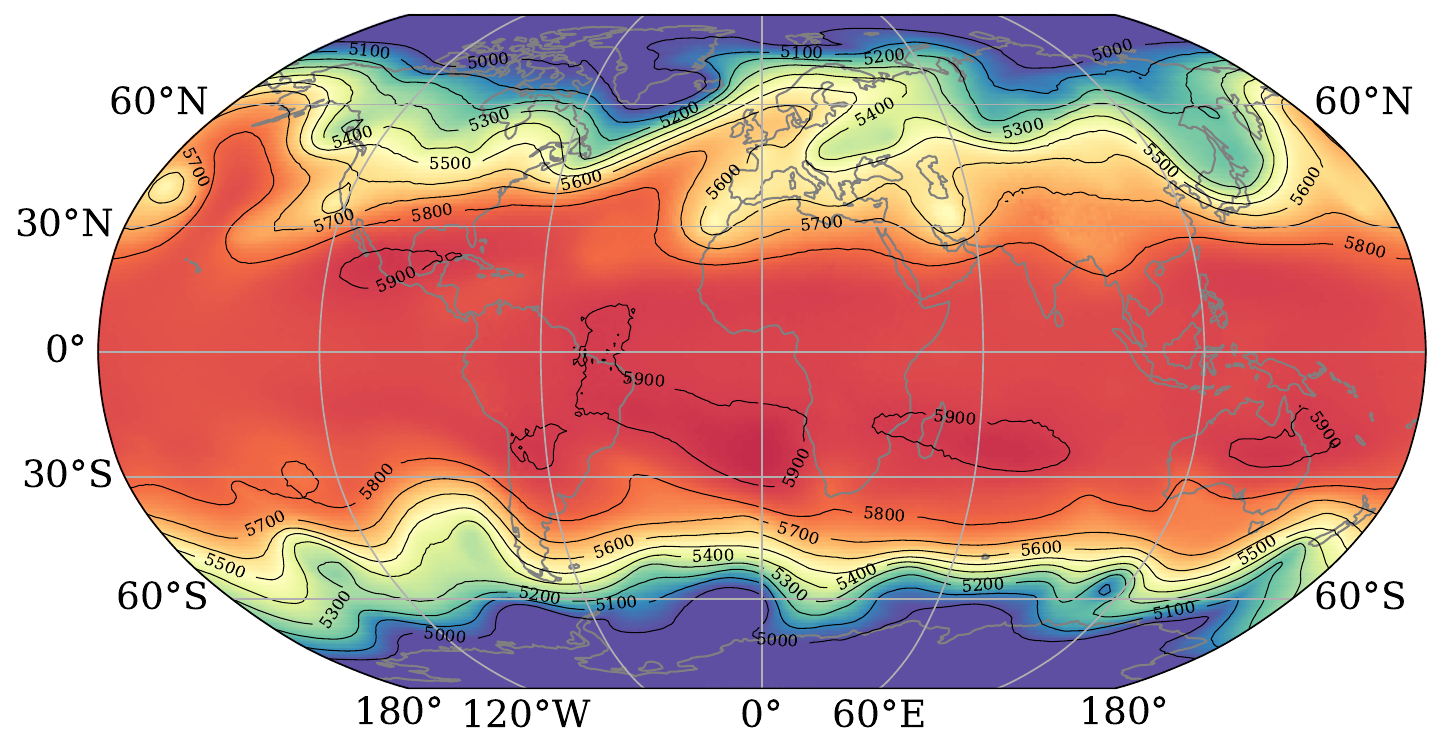}{10-day optimized forecast}
  \panel{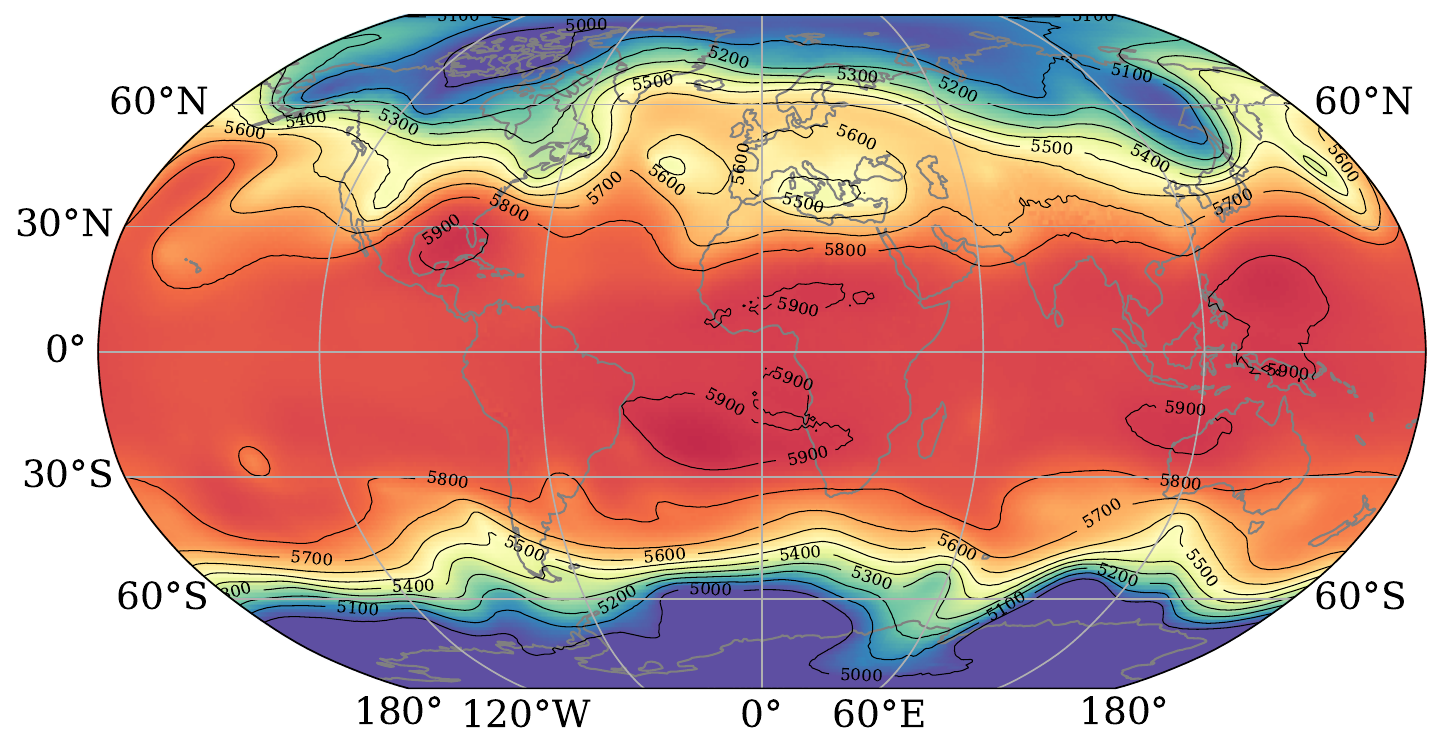}{15-day optimized forecast}
  \panel{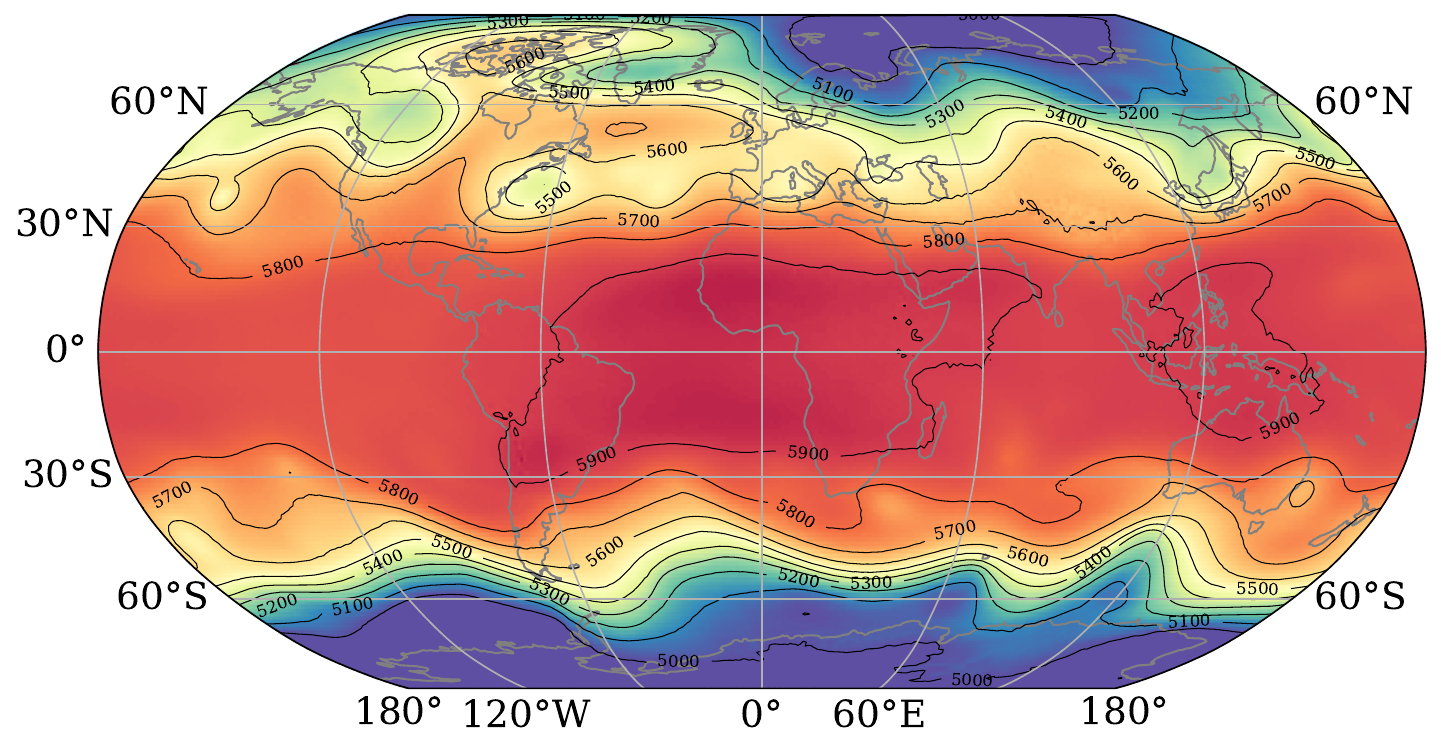}{20-day optimized forecast}
  \panel{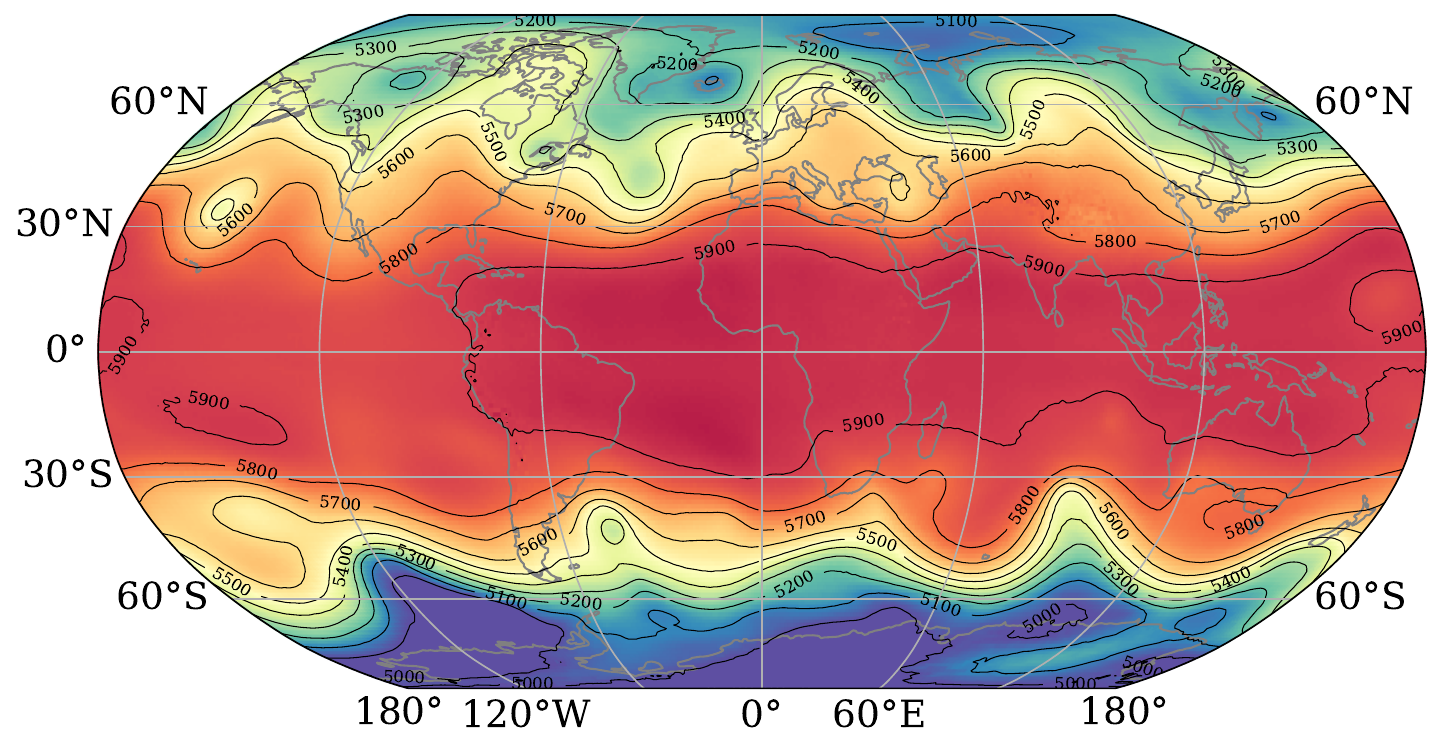}{25-day optimized forecast}
  \includegraphics[width=\linewidth,height=\panelH,keepaspectratio]{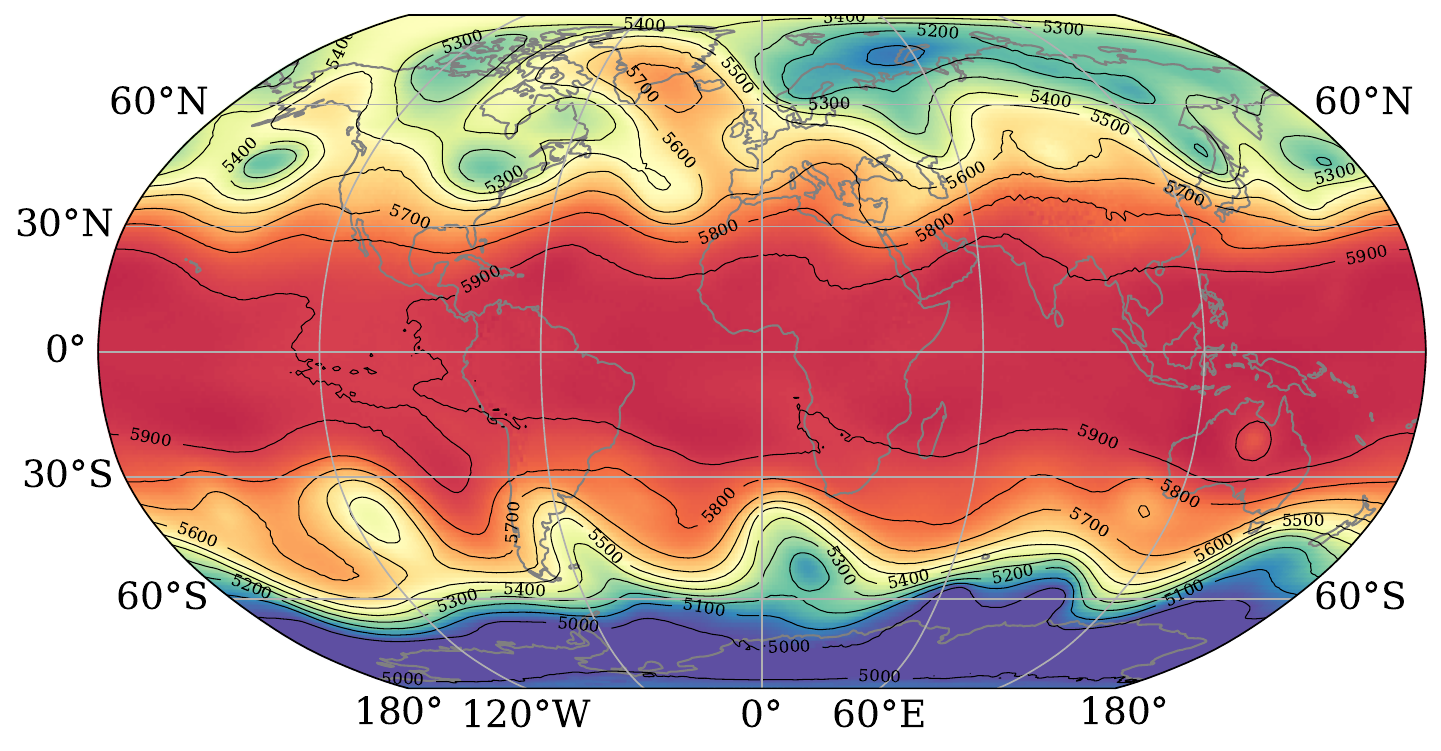}\par
  {\scriptsize 30-day optimized forecast}
\end{minipage}\hfill
\begin{minipage}[t]{0.32\textwidth}\centering
  \panel{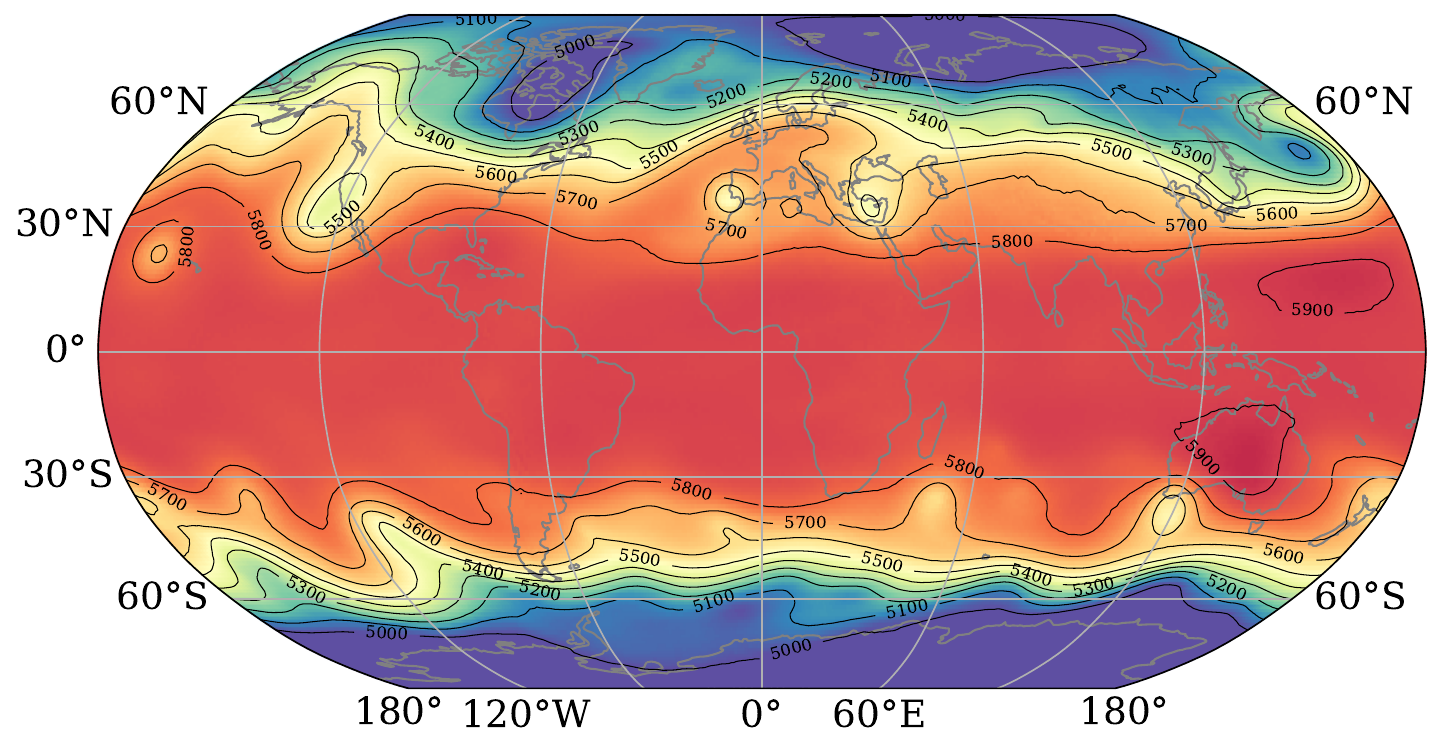}{5-day control forecast}
  \panel{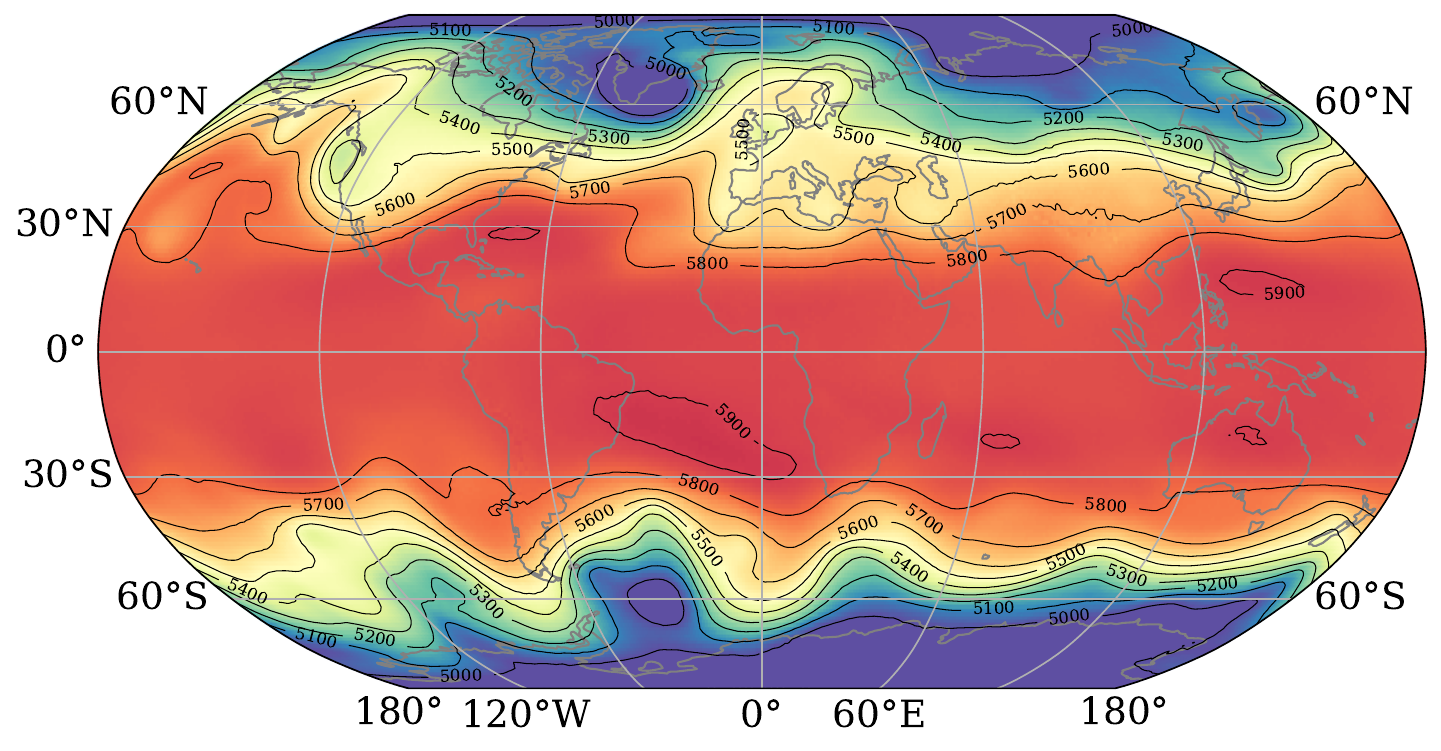}{10-day control forecast}
  \panel{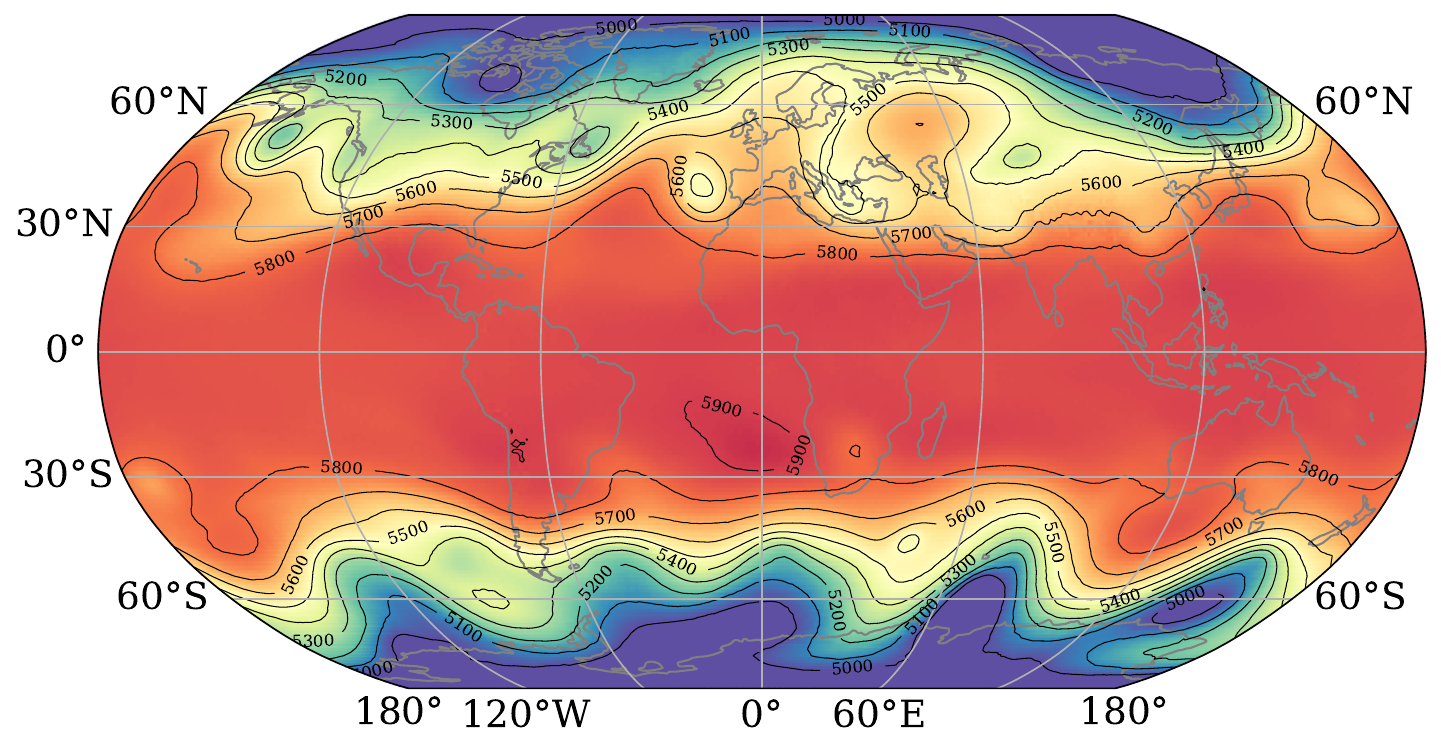}{15-day control forecast}
  \panel{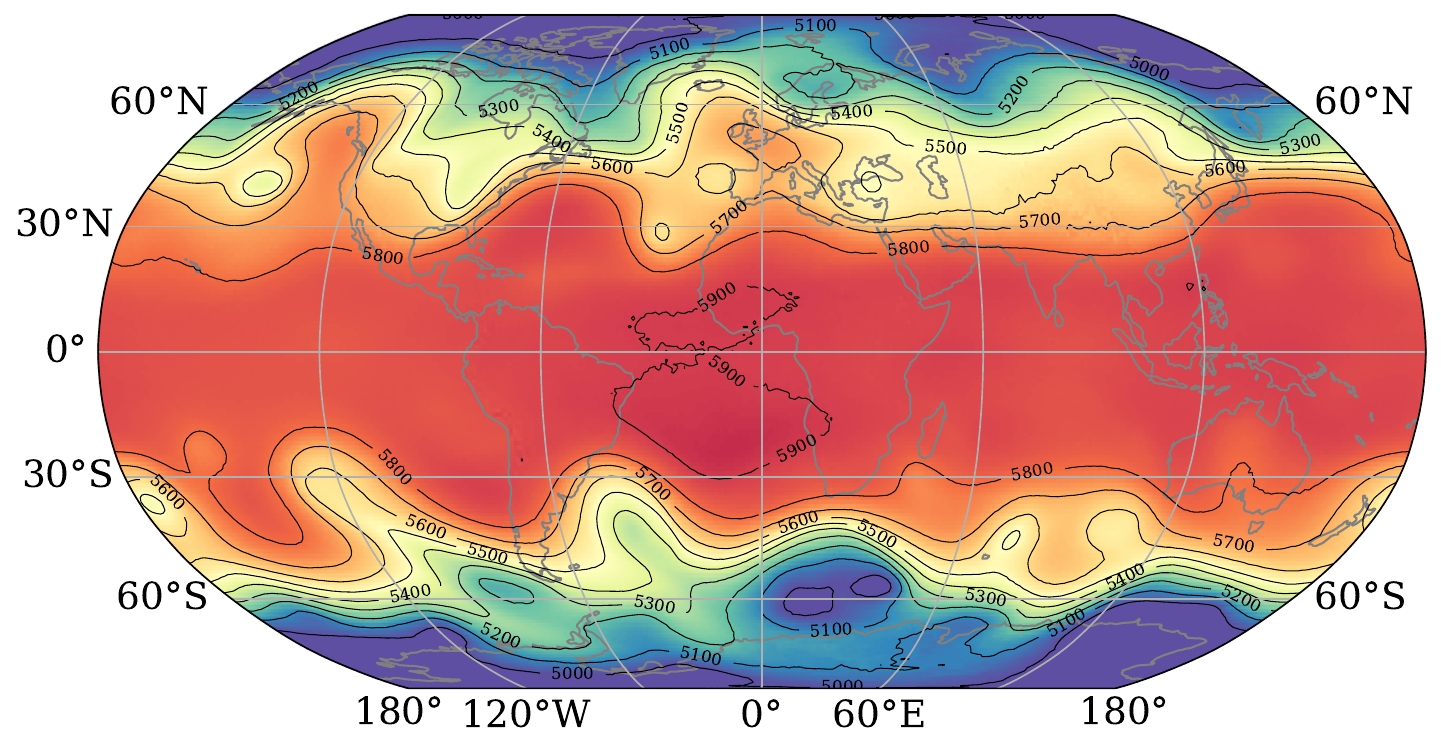}{20-day control forecast}
  \panel{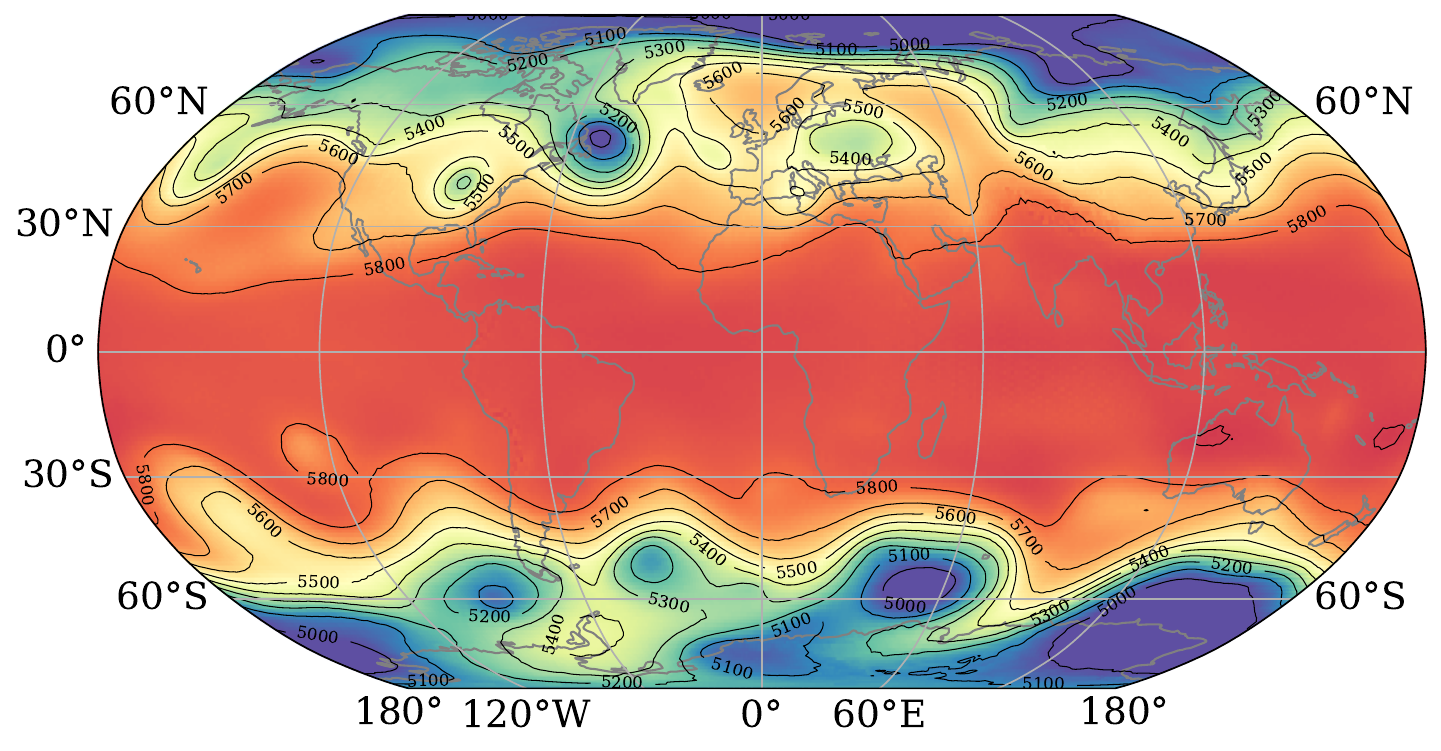}{25-day control forecast}
  \includegraphics[width=\linewidth,height=\panelH,keepaspectratio]{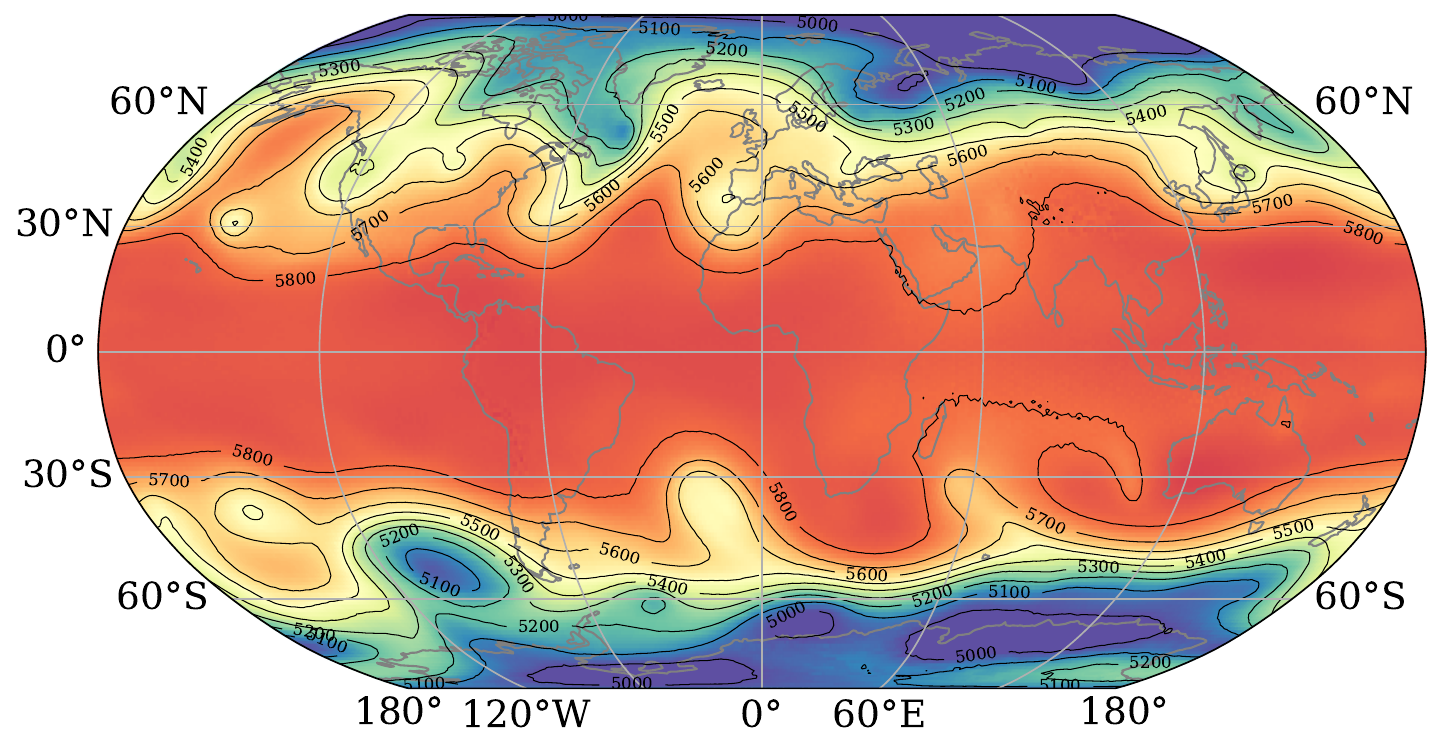}\par
  {\scriptsize 30-day control forecast}
\end{minipage}

\renewcommand{\thefigure}{B6}
\caption{The 500hPa geopotential height fields (Z500) for the worst-performing 30-day optimal forecast (defined as the largest time-integrated Z500 mean squared error) and its corresponding control, shown at 5, 10, 15, 20, 25, and 30-day lead times. Forecasts are initiated at 00Z 13 March 2020. The color bar in the upper-left ERA5 panel applies to all plots. Despite being the worst 30-day case among the 60 double-precision optimizations, the optimal forecast maintains high visual fidelity through day 20, capturing the strong ridge over Canada and the trough over Korea.}
\label{fig_s8:z500_plots}
\end{figure}

\clearpage
\begin{table}[t]
\centering
\setlength{\tabcolsep}{20pt}
\footnotesize
\begin{tabular}{@{}lccc@{}}
\hline
\textbf{Variable} 
  & \textbf{Mean Pert.\ Mag.} 
  & \textbf{Mean St.\ Dev.} 
  & \textbf{Grid Max St.\ Dev.} \\
\hline
200 hPa Zonal Wind              
  & 0.04 m s$^{-1}$      
  & 0.39 m s$^{-1}$      
  & 0.88 m s$^{-1}$       \\
200 hPa Meridional Wind         
  & 0.03 m s$^{-1}$      
  & 0.28 m s$^{-1}$      
  & 0.62 m s$^{-1}$       \\
500 hPa Geopotential Height     
  & 0.63 m               
  & 5.0 m                
  & 10.6 m                \\
500 hPa Pressure Vertical Velocity 
  & $8\times10^{-4}$ Pa s$^{-1}$ 
  & $5\times10^{-3}$ Pa s$^{-1}$ 
  & $1.1\times10^{-2}$ Pa s$^{-1}$ \\
700 hPa Specific Humidity       
  & 0.02 g kg$^{-1}$     
  & 0.08 g kg$^{-1}$     
  & 0.16 g kg$^{-1}$      \\
850 hPa Temperature             
  & 0.04 K               
  & 0.33 K               
  & 0.68 K                \\
\hline
\end{tabular}
\normalsize
\caption{Mean absolute value, mean standard deviation, and grid maximum standard deviation for select GraphCast upper-air variables for the optimal perturbations (single-precision optimizations).}
\label{table1}
\end{table}
\clearpage

\bibliographystyle{plainnat}   
\bibliography{references}

@STRING{AN        = "Astrophys.\ Norv."}

@STRING{CC        = "Climate Change"}

@STRING{CHAOS     = "Chaos"}

@STRING{MA        = "Meteor.\ Appl."}

@STRING{TELLUS    = "Tellus"}

@article{palmer2014real,
  title={The real butterfly effect},
  author={Palmer, Tim N and D{\"o}ring, Andreas and Seregin, Gregory},
  journal={Nonlinearity},
  volume={27},
  number={9},
  pages={R123},
  year={2014},
  publisher={IOP Publishing}
}

@article{Swanson-1998,
 author = {Swanson, Kyle and Vautard, Robert and Pires, Carlos},
 doi = {10.3402/tellusa.v50i4.14540},
 journal = {Tellus A: Dynamic Meteorology and Oceanography},
 month = {Jan},
 title = {Four-dimensional variational assimilation and predictability in a quasi-geostrophic model},
 year = {1998}
}

@misc{weatherbenchx2025,
  author = {{Google Research}},
  title = {WeatherBench-X: A modular framework for evaluating weather forecasts},
  year = {2025},
  publisher = {GitHub},
  howpublished = {\url{https://github.com/google-research/weatherbenchX}},
  note = {Apache License, Version 2.0}
}

@book{Wilks2011,
  author    = {Daniel S. Wilks},
  title     = {Statistical Methods in the Atmospheric Sciences},
  edition   = {3},
  publisher = {Academic Press},
  year      = {2011},
  isbn      = {9780123850225}
}

@article{McLay2022,
  author  = {McLay, Justin G. and Satterfield, Elizabeth},
  title   = {Forecast Dropouts in the {NAVGEM} Model: Characterization with Respect to Other Models, Large-Scale Indices, and Ensemble Forecasts},
  journal = {Weather and Forecasting},
  year    = {2022},
  volume  = {37},
  number  = {11},
  pages   = {1945--1964},
  doi     = {10.1175/WAF-D-21-0208.1}
}

@article{Lillo2017,
  author  = {Lillo, Samuel P. and Parsons, David B.},
  title   = {Investigating the Dynamics of Error Growth in {ECMWF} Medium-Range Forecast Busts},
  journal = {Quarterly Journal of the Royal Meteorological Society},
  year    = {2017},
  volume  = {143},
  number  = {704},
  pages   = {1211--1226},
  doi     = {10.1002/qj.2938}
}

@article{rodwell2013,
  title={Characteristics of occasional poor medium-range weather forecasts for Europe},
  author={Rodwell, Mark J and Magnusson, Linus and Bauer, Peter and Bechtold, Peter and Bonavita, Massimo and Cardinali, Carla and Diamantakis, Michail and Earnshaw, Paul and Garcia-Mendez, Antonio and Isaksen, Lars and others},
  journal={Bulletin of the American Meteorological Society},
  volume={94},
  number={9},
  pages={1393--1405},
  year={2013},
  publisher={American Meteorological Society}
}

@article{pena2014,
  author    = {Peña, Marcelo and Toth, Zoltan},
  title     = {Estimation of analysis and forecast error variances},
  journal   = {Tellus A: Dynamic Meteorology and Oceanography},
  volume    = {66},
  number    = {1},
  pages     = {21767},
  year      = {2014},
  doi       = {10.3402/tellusa.v66.21767},
  url       = {https://doi.org/10.3402/tellusa.v66.21767}
}

@article { hakim2005,
      author = "Gregory J. Hakim",
      title = "Vertical Structure of Midlatitude Analysis and Forecast Errors",
      journal = "Monthly Weather Review",
      year = "2005",
      publisher = "American Meteorological Society",
      address = "Boston MA, USA",
      volume = "133",
      number = "3",
      doi = "10.1175/MWR-2882.1",
      pages=      "567 - 578",
      url = "https://journals.ametsoc.org/view/journals/mwre/133/3/mwr-2882.1.xml"
}

@article {analysis1986,
      author = "Roger  Daley and Thomas  Mayer",
      title = "Estimates of Global Analysis Error from the Global Weather Experiment Observational Network",
      journal = "Monthly Weather Review",
      year = "1986",
      publisher = "American Meteorological Society",
      address = "Boston MA, USA",
      volume = "114",
      number = "9",
      doi = "10.1175/1520-0493(1986)114<1642:EOGAEF>2.0.CO;2",
      pages=      "1642 - 1653",
      url = "https://journals.ametsoc.org/view/journals/mwre/114/9/1520-0493_1986_114_1642_eogaef_2_0_co_2.xml"
}

@article{Pires1996,
  title={On extending the limits of variational assimilation in nonlinear chaotic systems},
  author={Carlos A. L. Pires and Robert Vautard and O. Talagrand},
  journal={Tellus A},
  year={1996},
  volume={48},
  pages={96-121},
  url={https://api.semanticscholar.org/CorpusID:122300156}
}

@article{heatwave_2024,
author = {Vonich, P. Trent and Hakim, Gregory J.},
title = {Predictability Limit of the 2021 Pacific Northwest Heatwave From Deep-Learning Sensitivity Analysis},
journal = {Geophysical Research Letters},
volume = {51},
year={2024},
number = {19},
pages = {e2024GL110651},
doi = {https://doi.org/10.1029/2024GL110651},
url = {https://agupubs.onlinelibrary.wiley.com/doi/abs/10.1029/2024GL110651},
eprint = {https://agupubs.onlinelibrary.wiley.com/doi/pdf/10.1029/2024GL110651},
note = {e2024GL110651 2024GL110651}}

@article{bi2023,
  title={Accurate medium-range global weather forecasting with {{3D}} neural networks},
  author={Bi, Kaifeng and Xie, Lingxi and Zhang, Hengheng and Chen, Xin and Gu, Xiaotao and Tian, Qi},
  journal={Nature},
  pages={1--6},
  year={2023},
  howpublished = {[Software]},
  publisher={Nature Publishing Group UK London}
}

@article{rasp2023weatherbench,
  title={Weatherbench 2: A benchmark for the next generation of data-driven global weather models},
  author={Rasp, Stephan and Hoyer, Stephan and Merose, Alexander and Langmore, Ian and Battaglia, Peter and Russel, Tyler and Sanchez-Gonzalez, Alvaro and Yang, Vivian and Carver, Rob and Agrawal, Shreya and others},
  journal={arXiv preprint arXiv:2308.15560},
  year={2023}
}

@article{lam2023learning,
  title={Learning skillful medium-range global weather forecasting},
  author={Lam, Remi and Sanchez-Gonzalez, Alvaro and Willson, Matthew and Wirnsberger, Peter and Fortunato, Meire and Alet, Ferran and Ravuri, Suman and Ewalds, Timo and Eaton-Rosen, Zach and Hu, Weihua and others},
  journal={Science},
  volume={382},
  number={6677},
  pages={1416--1421},
  year={2023},
  howpublished = {[Software]},
  publisher={American Association for the Advancement of Science}
}

@article{hersbach2020era5,
  title={The ERA5 global reanalysis},
  author={Hersbach, Hans and Bell, Bill and Berrisford, Paul and Hirahara, Shoji and Hor{\'a}nyi, Andr{\'a}s and Mu{\~n}oz-Sabater, Joaqu{\'\i}n and Nicolas, Julien and Peubey, Carole and Radu, Raluca and Schepers, Dinand and others},
  journal={Quarterly Journal of the Royal Meteorological Society},
  volume={146},
  number={730},
  pages={1999--2049},
  year={2020},
  publisher={Wiley Online Library}
}

@article{thompson20222021,
  title={The 2021 western North America heat wave among the most extreme events ever recorded globally},
  author={Thompson, Vikki and Kennedy-Asser, Alan T and Vosper, Emily and Lo, YT Eunice and Huntingford, Chris and Andrews, Oliver and Collins, Matthew and Hegerl, Gabrielle C and Mitchell, Dann},
  journal={Science Advances},
  volume={8},
  number={18},
  pages={eabm6860},
  year={2022},
  publisher={American Association for the Advancement of Science}
}

@article{leach2024heatwave,
  title={Heatwave attribution based on reliable operational weather forecasts},
  author={Leach, Nicholas J and Roberts, Christopher D and Aengenheyster, Matthias and Heathcote, Daniel and Mitchell, Dann M and Thompson, Vikki and Palmer, Tim and Weisheimer, Antje and Allen, Myles R},
  journal={Nature Communications},
  volume={15},
  number={1},
  pages={4530},
  year={2024},
  publisher={Nature Publishing Group UK London}
}

@article{buizza2015forecast,
  title={The forecast skill horizon},
  author={Buizza, Roberto and Leutbecher, Martin},
  journal={Quarterly Journal of the Royal Meteorological Society},
  volume={141},
  number={693},
  pages={3366--3382},
  year={2015},
  publisher={Wiley Online Library}
}

@article{Lorenz1969,
  author = {Lorenz, Edward N.},
  title = {The Predictability of a Flow Which Possesses Many Scales of Motion},
  journal = {Tellus},
  volume = {21},
  number = {3},
  pages = {289--307},
  year = {1969},
  doi = {10.1111/j.2153-3490.1969.tb00444.x},
  url = {http://eaps4.mit.edu/research/Lorenz/publications.htm}
}

@article{merra2,
  author    = {Gelaro, Ronald and McCarty, Will and Su{\'a}rez, Max J. and Todling, Ricardo and Molod, Andrea and Takacs, Lawrence and Randles, Cynthia A. and Darmenov, Anton and Bosilovich, Michael G. and Reichle, Rolf and others},
  title     = {The Modern-Era Retrospective Analysis for Research and Applications, Version 2 (MERRA-2)},
  journal   = {Journal of Climate},
  volume    = {30},
  number    = {14},
  pages     = {5419--5454},
  year      = {2017},
  doi       = {10.1175/JCLI-D-16-0758.1}
}

@article{LA_NINA_2020,
  author = {Li, Xiaofan and Hu, Zeng-Zhen and Tseng, Yu-heng and Liu, Yunyun and Liang, Ping},
  title = {A Historical Perspective of the La Niña Event in 2020/2021},
  journal = {Journal of Geophysical Research: Atmospheres},
  volume = {127},
  number = {7},
  year = {2022},
      doi = {10.1029/2021jd035546}}

@article {palmer1988,
      author = "T. N.  Palmer and S.  Tibaldi",
      title = "On the Prediction of Forecast Skill",
      journal = "Monthly Weather Review",
      year = "1988",
      publisher = "American Meteorological Society",
      address = "Boston MA, USA",
      volume = "116",
      number = "12",
      doi = "10.1175/1520-0493(1988)116<2453:OTPOFS>2.0.CO;2",
      pages=      "2453 - 2480",
      url = "https://journals.ametsoc.org/view/journals/mwre/116/12/1520-0493_1988_116_2453_otpofs_2_0_co_2.xml"
}

@article {dalcher1987,
      author = "Eugenia  Kalnay and Amnon  Dalcher",
      title = "Forecasting Forecast Skill",
      journal = "Monthly Weather Review",
      year = "1987",
      publisher = "American Meteorological Society",
      address = "Boston MA, USA",
      volume = "115",
      number = "2",
      doi = "10.1175/1520-0493(1987)115<0349:FFS>2.0.CO;2",
      pages=      "349 - 356",
      url = "https://journals.ametsoc.org/view/journals/mwre/115/2/1520-0493_1987_115_0349_ffs_2_0_co_2.xml"
}

@article{ie_ne,
author = {Bonavita, Massimo and Geer, Alan J.},
title = {Forecast verification using information and noise},
journal = {Quarterly Journal of the Royal Meteorological Society},
year = "2026",
pages = {e70109},
doi = {https://doi.org/10.1002/qj.70109},
url = {https://rmets.onlinelibrary.wiley.com/doi/abs/10.1002/qj.70109},
eprint = {https://rmets.onlinelibrary.wiley.com/doi/pdf/10.1002/qj.70109},
note = {e70109 QJ-25-0227.R1},
}

@article {judt2020,
      author = "Falko Judt",
      title = "Atmospheric Predictability of the Tropics, Middle Latitudes, and Polar Regions Explored through Global Storm-Resolving Simulations",
      journal = "Journal of the Atmospheric Sciences",
      year = "2020",
      publisher = "American Meteorological Society",
      address = "Boston MA, USA",
      volume = "77",
      number = "1",
      doi = "10.1175/JAS-D-19-0116.1",
      pages=      "257 - 276",
      url = "https://journals.ametsoc.org/view/journals/atsc/77/1/jas-d-19-0116.1.xml"
}

@article { judt2018,
      author = "Falko Judt",
      title = "Insights into Atmospheric Predictability through Global Convection-Permitting Model Simulations",
      journal = "Journal of the Atmospheric Sciences",
      year = "2018",
      publisher = "American Meteorological Society",
      address = "Boston MA, USA",
      volume = "75",
      number = "5",
      doi = "10.1175/JAS-D-17-0343.1",
      pages=      "1477 - 1497",
      url = "https://journals.ametsoc.org/view/journals/atsc/75/5/jas-d-17-0343.1.xml"
}

@article {selz2019,
      author = "Tobias Selz",
      title = "Estimating the Intrinsic Limit of Predictability Using a Stochastic Convection Scheme",
      journal = "Journal of the Atmospheric Sciences",
      year = "2019",
      publisher = "American Meteorological Society",
      address = "Boston MA, USA",
      volume = "76",
      number = "3",
      doi = "10.1175/JAS-D-17-0373.1",
      pages=      "757 - 765",
      url = "https://journals.ametsoc.org/view/journals/atsc/76/3/jas-d-17-0373.1.xml"
}

@misc{cresswell2025_earth,
      title={A Deep Learning Earth System Model for Efficient Simulation of the Observed Climate}, 
      author={Nathaniel Cresswell-Clay and Bowen Liu and Dale Durran and Zihui Liu and Zachary I. Espinosa and Raul Moreno and Matthias Karlbauer},
      year={2025},
      eprint={2409.16247},
      archivePrefix={arXiv},
      primaryClass={physics.ao-ph},
      url={https://arxiv.org/abs/2409.16247}, 
}

@article {lloveras2025_goose,
      author = "Daniel J. Lloveras and James D. Doyle and Dale R. Durran",
      title = "Can Observation Targeting Be a Wild Goose Chase? An Adjoint-Sensitivity Study of a U.S. East Coast Cyclone Forecast Bust",
      journal = "Journal of the Atmospheric Sciences",
      year = "2025",
      publisher = "American Meteorological Society",
      address = "Boston MA, USA",
      volume = "82",
      number = "2",
      doi = "10.1175/JAS-D-24-0044.1",
      pages=      "343 - 360",
      url = "https://journals.ametsoc.org/view/journals/atsc/82/2/JAS-D-24-0044.1.xml"
}

@article{zhang2019predictability,
  title={What is the predictability limit of midlatitude weather?},
  author={Zhang, Fuqing and Sun, Y Qiang and Magnusson, Linus and Buizza, Roberto and Lin, Shian-Jiann and Chen, Jan-Huey and Emanuel, Kerry},
  journal={Journal of the Atmospheric Sciences},
  volume={76},
  number={4},
  pages={1077--1091},
  year={2019},
  publisher={American Meteorological Society}
}

@inproceedings{lorenz1996predictability,
  title={Predictability: {{A}} problem partly solved},
  author={Lorenz, Edward N},
  booktitle={Proc. Seminar on predictability},
  volume={1},
  year={1996},
  organization={Reading}
}

@article{errico1997adjoint,
  title={What is an adjoint model?},
  author={Errico, Ronald M},
  journal={Bulletin of the American Meteorological Society},
  volume={78},
  number={11},
  pages={2577--2592},
  year={1997},
  publisher={American Meteorological Society}
}

@article{langland1995,
  title={Evaluation of physical processes in an idealized extratropical cyclone using adjoint sensitivity},
  author={Langland, Rolf H and Elsberry, Russell L and Errico, Ronald M},
  journal={Quarterly Journal of the Royal Meteorological Society},
  volume={121},
  number={526},
  pages={1349--1386},
  year={1995},
  publisher={Wiley Online Library}
}

@article { langland2002,
      author = "Rolf H. Langland and Melvyn A. Shapiro and Ronald Gelaro",
      title = "Initial Condition Sensitivity and Error Growth in Forecasts of the 25 January 2000 East Coast Snowstorm",
      journal = "Monthly Weather Review",
      year = "2002",
      publisher = "American Meteorological Society",
      address = "Boston MA, USA",
      volume = "130",
      number = "4",
      doi = "10.1175/1520-0493(2002)130<0957:ICSAEG>2.0.CO;2",
      pages=      "957 - 974",
      url = "https://journals.ametsoc.org/view/journals/mwre/130/4/1520-0493_2002_130_0957_icsaeg_2.0.co_2.xml"
}

@article{aurora,
  title   = {A foundation model for the Earth system},
  author  = {Bodnar, Cristian and Bruinsma, Wessel P. and Lucic, Ana and Stanley, Megan and Allen, Anna and Brandstetter, Johannes and Garvan, Patrick and Riechert, Maik and Weyn, Jonathan A. and Dong, Haiyu and Gupta, Jayesh K. and Thambiratnam, Kit and Archibald, Alexander T. and Wu, Chun-Chieh and Heider, Elizabeth and Welling, Max and Turner, Richard E. and Perdikaris, Paris},
  journal = {Nature},
  year    = {2025},
  volume  = {641},
  number  = {8065},
  pages   = {1180--1187},
  doi     = {10.1038/s41586-025-09005-y},
  url     = {https://doi.org/10.1038/s41586-025-09005-y},
  isbn    = {1476-4687},
}

@misc{ecmwf_activity,
  author       = "{Zied Ben Bouallègue and the AIFS team}",
  title        = "{Accuracy versus activity}",
  year         = {2024},
  month        = dec,
  howpublished = "\url{https://www.ecmwf.int/en/about/media-centre/aifs-blog/2024/accuracy-versus-activity}",
  note         = "[Online; accessed 5-September-2025]"
}

@article{selz2023,
  title={Can Artificial Intelligence-Based Weather Prediction Models Simulate the Butterfly Effect?},
  author={Selz, Tobias and Craig, George C},
  journal={Geophysical Research Letters},
  volume={50},
  number={20},
  pages={e2023GL105747},
  year={2023},
  publisher={Wiley Online Library}
}

@article{bonavita2024,
author = {Bonavita, Massimo},
title = {On Some Limitations of Current Machine Learning Weather Prediction Models},
journal = {Geophysical Research Letters},
volume = {51},
number = {12},
pages = {e2023GL107377},
doi = {https://doi.org/10.1029/2023GL107377},
url = {https://agupubs.onlinelibrary.wiley.com/doi/abs/10.1029/2023GL107377},
eprint = {https://agupubs.onlinelibrary.wiley.com/doi/pdf/10.1029/2023GL107377},
note = {e2023GL107377 2023GL107377},
year = {2024}
}

@article{Shen2022,
  author    = {Shen, Bo-Wen and Pielke, Roger and Zeng, Xubin and Cui, Jialin and Faghih-Naini, Sara and Paxson, Wei and Kesarkar, Amit and Zeng, Xiping and Atlas, Robert},
  title     = {The Dual Nature of Chaos and Order in the Atmosphere},
  journal   = {Atmosphere},
  volume    = {13},
  number    = {11},
  year      = {2022},
  pages     = {1892},
  doi       = {10.3390/atmos13111892},
  url       = {https://www.mdpi.com/2073-4433/13/11/1892},
  issn      = {2073-4433}
}

@misc{adv_feature,
      title={Adversarial Examples Are Not Bugs, They Are Features}, 
      author={Andrew Ilyas and Shibani Santurkar and Dimitris Tsipras and Logan Engstrom and Brandon Tran and Aleksander Madry},
      year={2019},
      eprint={1905.02175},
      archivePrefix={arXiv},
      primaryClass={stat.ML},
      url={https://arxiv.org/abs/1905.02175}, 
}

@article{not_so_weird,
  title={Adversarial Perturbations Are Not So Weird: Entanglement of Robust and Non-Robust Features in Neural Network Classifiers},
  author={Jacob Mitchell Springer and Melanie Mitchell and Garrett T. Kenyon},
  journal={ArXiv},
  year={2021},
  volume={abs/2102.05110},
  url={https://api.semanticscholar.org/CorpusID:231861678}
}

@article{adv_space,
  author       = {Pedro Tabacof and
                  Eduardo Valle},
  title        = {Exploring the Space of Adversarial Images},
  journal      = {CoRR},
  volume       = {abs/1510.05328},
  year         = {2015},
  url          = {http://arxiv.org/abs/1510.05328},
  eprinttype    = {arXiv},
  eprint       = {1510.05328},
  bibsource    = {dblp computer science bibliography, https://dblp.org}
}

@article{Shen2023,
  author    = {Shen, Bo-Wen and Pielke, Roger A. and Zeng, Xubin and Zeng, Xiping},
  title     = {Lorenz's View on the Predictability Limit of the Atmosphere},
  journal   = {Encyclopedia},
  volume    = {3},
  number    = {3},
  year      = {2023},
  pages     = {887--899},
  doi       = {10.3390/encyclopedia3030063},
  url       = {https://www.mdpi.com/2673-8392/3/3/63},
  issn      = {2673-8392}}

@misc{szegedy2014,
      title={Intriguing properties of neural networks}, 
      author={Christian Szegedy and Wojciech Zaremba and Ilya Sutskever and Joan Bruna and Dumitru Erhan and Ian Goodfellow and Rob Fergus},
      year={2014},
      eprint={1312.6199},
      archivePrefix={arXiv},
      primaryClass={cs.CV},
      url={https://arxiv.org/abs/1312.6199}, 
}

@misc{adversarialexamples,
      title={Explaining and Harnessing Adversarial Examples}, 
      author={Ian J. Goodfellow and Jonathon Shlens and Christian Szegedy},
      year={2015},
      eprint={1412.6572},
      archivePrefix={arXiv},
      primaryClass={stat.ML},
      url={https://arxiv.org/abs/1412.6572}, 
}

@article{deepfool,
author = {Moosavi-Dezfooli, Seyed-Mohsen and Fawzi, Alhussein and Frossard, Pascal},
year = {2016},
month = {11},
pages = {},
title = {DeepFool: a simple and accurate method to fool deep neural networks},
journal = {CVPR}
}

@article{healpix2024,
author = {Karlbauer, Matthias and Cresswell-Clay, Nathaniel and Durran, Dale R. and Moreno, Raul A. and Kurth, Thorsten and Bonev, Boris and Brenowitz, Noah and Butz, Martin V.},
title = {Advancing Parsimonious Deep Learning Weather Prediction Using the HEALPix Mesh},
journal = {Journal of Advances in Modeling Earth Systems},
volume = {16},
number = {8},
pages = {e2023MS004021},
doi = {https://doi.org/10.1029/2023MS004021},
url = {https://agupubs.onlinelibrary.wiley.com/doi/abs/10.1029/2023MS004021},
eprint = {https://agupubs.onlinelibrary.wiley.com/doi/pdf/10.1029/2023MS004021},
note = {e2023MS004021 2023MS004021},
year = {2024}
}

@article{Shen2024,
  author = {Shen, Bo-Wen and Pielke, Sr., Roger A. and Zeng, Xubin and Zeng, Xiping},
  title = {Exploring the Origin of the Two-Week Predictability Limit: A Revisit of Lorenz’s Predictability Studies in the 1960s},
  journal = {Atmosphere},
  volume = {15},
  number = {7},
  pages = {837},
  year = {2024},
  doi = {10.3390/atmos15070837},
  url = {https://doi.org/10.3390/atmos15070837}
}

@article { doyle2012,
      author = "James D. Doyle and Carolyn A. Reynolds and Clark Amerault and Jonathan Moskaitis",
      title = "Adjoint Sensitivity and Predictability of Tropical Cyclogenesis",
      journal = "Journal of the Atmospheric Sciences",
      year = "2012",
      publisher = "American Meteorological Society",
      address = "Boston MA, USA",
      volume = "69",
      number = "12",
      doi = "10.1175/JAS-D-12-0110.1",
      pages=      "3535 - 3557",
      url = "https://journals.ametsoc.org/view/journals/atsc/69/12/jas-d-12-0110.1.xml"
}

@article { doyle2014,
      author = "James D. Doyle and Clark Amerault and Carolyn A. Reynolds and P. Alex Reinecke",
      title = "Initial Condition Sensitivity and Predictability of a Severe Extratropical Cyclone Using a Moist Adjoint",
      journal = "Monthly Weather Review",
      year = "2014",
      publisher = "American Meteorological Society",
      address = "Boston MA, USA",
      volume = "142",
      number = "1",
      doi = "10.1175/MWR-D-13-00201.1",
      pages=      "320 - 342",
      url = "https://journals.ametsoc.org/view/journals/mwre/142/1/mwr-d-13-00201.1.xml"
}

@article{doyle2019,
  title={Adjoint sensitivity analysis of high-impact extratropical cyclones},
  author={Doyle, James D and Reynolds, Carolyn A and Amerault, Clark},
  journal={Monthly Weather Review},
  volume={147},
  number={12},
  pages={4511--4532},
  year={2019}
}

@misc{jax2018github,
  author = {James Bradbury and Roy Frostig and Peter Hawkins and Matthew James Johnson and Chris Leary and Dougal Maclaurin and George Necula and Adam Paszke and Jake Vander{P}las and Skye Wanderman-{M}ilne and Qiao Zhang},
  title = {{JAX}: composable transformations of {P}ython+{N}um{P}y programs},
  url = {http://github.com/google/jax},
  version = {0.3.13},
  year = {2018},
}

@article{Li_2024,
  author = {Zongheng Li and Jun Peng and Lifeng Zhang and Jiping Guan},
  title = {Exploring the Differences in Kinetic Energy Spectra between the NCEP FNL and ERA5 Datasets},
  journal = {Journal of the Atmospheric Sciences},
  volume = {81},
  number = {2},
  pages = {363--380},
  year = {2024},
  month = {February},
  doi = {10.1175/JAS-D-23-0043.1},
  publisher = {American Meteorological Society},
  url = {https://doi.org/10.1175/JAS-D-23-0043.1}
}

@misc{kingma2017,
      title={Adam: A Method for Stochastic Optimization}, 
      author={Diederik P. Kingma and Jimmy Ba},
      year={2017},
      eprint={1412.6980},
      archivePrefix={arXiv},
      primaryClass={cs.LG},
      url={https://arxiv.org/abs/1412.6980}, 
}

@article{bano_medina2025,
  author = {Baño-Medina, Jorge and Sengupta, Agniv and Doyle, James D. and Reynolds, Carolyn A. and Watson-Parris, Duncan and Monache, Luca Delle},
  title = {Are AI weather models learning atmospheric physics? A sensitivity analysis of cyclone Xynthia},
  journal = {npj Climate and Atmospheric Science},
  year = {2025},
  volume = {8},
  number = {1},
  pages = {92},

  doi = {10.1038/s41612-025-00949-6},
  isbn = {2397-3722},
  url = {https://doi.org/10.1038/s41612-025-00949-6},
  date = {2025/03/07},
}

@misc{hersbach2017era5,
  author = {Hersbach, H. and Bell, B. and Berrisford, P. and Hirahara, S. and Horányi, A. and Muñoz‐Sabater, J. and Nicolas, J. and Peubey, C. and Radu, R. and Schepers, D. and Simmons, A. and Soci, C. and Abdalla, S. and Abellan, X. and Balsamo, G. and Bechtold, P. and Biavati, G. and Bidlot, J. and Bonavita, M. and De Chiara, G. and Dahlgren, P. and Dee, D. and Diamantakis, M. and Dragani, R. and Flemming, J. and Forbes, R. and Fuentes, M. and Geer, A. and Haimberger, L. and Healy, S. and Hogan, R.J. and Hólm, E. and Janisková, M. and Keeley, S. and Laloyaux, P. and Lopez, P. and Lupu, C. and Radnoti, G. and de Rosnay, P. and Rozum, I. and Vamborg, F. and Villaume, S. and Thépaut, J-N.},
  title = {Complete ERA5 from 1940: Fifth generation of ECMWF atmospheric reanalyses of the global climate},
  year = {2017},
  howpublished = {[Dataset]. Copernicus Climate Change Service (C3S) Data Store (CDS)},
  note = {DOI: 10.24381/cds.143582cf}
}

@misc{ziga2025,
  title         = {The {Changes} of the Northern {Hadley} {Cell} {Strength} in {Reanalyses} and {Radiosonde} {Observations}},
  author        = {Matic Pikovnik and {\v Z}iga Zaplotnik},
  year          = {2025},
  eprint        = {2503.05331},
  archivePrefix = {arXiv},
  primaryClass  = {physics.ao-ph},
  url           = {https://arxiv.org/abs/2503.05331}
}

@article{Charney1966,
  author = {Charney, J. G. and Fleagle, R. G. and Lally, V. E. and Riehl, H. and Wark, D. Q.},
  title = {The feasibility of a global observation and analysis experiment},
  journal = {Bulletin of the American Meteorological Society},
  year = {1966},
  volume = {47},
  pages = {200--220},
}

@article{brenowitz2024practical,
  author = {Brenowitz, Noah D. and Cohen, Yair and Pathak, Jaideep and Mahesh, Ankur and Bonev, Boris and Kurth, Thorsten and Durran, Dale R. and Harrington, Peter and Pritchard, Michael S.},
  title = {A Practical Probabilistic Benchmark for AI Weather Models},
  journal = {arXiv preprint arXiv:2401.15305},
  year = {2024},
  month = {jan},
  day = {27},
  volume = {2401.15305v1},
  archivePrefix = {arXiv},
  primaryClass = {physics.ao-ph},
  eprint = {2401.15305},
  url = {https://arxiv.org/abs/2401.15305},
  note = {License: CC BY 4.0},
 }

@article{Charlton-Perez2024,
  author = {Charlton-Perez, Andrew J. and Dacre, Helen F. and Driscoll, Simon and Gray, Suzanne L. and Harvey, Ben and Harvey, Natalie J. and Hunt, Kieran M. R. and Lee, Robert W. and Swaminathan, Ranjini and Vandaele, Remy and Volonté, Ambrogio},
  title = {Do AI models produce better weather forecasts than physics-based models? A quantitative evaluation case study of Storm Ciarán},
  journal = {npj Climate and Atmospheric Science},
  year = {2024},
  volume = {7},
  number = {1},
  pages = {93},
  month = {apr},
  day = {22},
  doi = {10.1038/s41612-024-00638-w},
  url = {https://doi.org/10.1038/s41612-024-00638-w},
  issn = {2397-3722},
}

\end{document}